\documentclass[lettersize,journal]{IEEEtran}
\usepackage{amsmath,amsfonts}
\usepackage{algorithmic}
\usepackage{array}
\usepackage[caption=false,font=normalsize,labelfont=sf,textfont=sf]{subfig}
\usepackage{textcomp}
\usepackage{stfloats}
\usepackage{url}
\usepackage{verbatim}
\usepackage{graphicx}
\def\BibTeX{{\rm B\kern-.05em{\sc i\kern-.025em b}\kern-.08em
    T\kern-.1667em\lower.7ex\hbox{E}\kern-.125emX}}
\usepackage{balance}

\usepackage{epsfig}
\usepackage{color,soul}
\RequirePackage{booktabs}
\usepackage{multirow}
\usepackage{caption}
\usepackage{makecell}
\usepackage{adjustbox}
\usepackage{rotating}
\usepackage{longtable}
\usepackage{tabularx}
\usepackage{lscape}

\newtheorem{definition}{Definition}

\begin{document}

\title{BELE: Blur Equivalent Linearized Estimator}

\author{Paolo Giannitrapani\thanks{Contact author: Paolo Giannitrapani, Dept. of Information Engineering, Electronics and Telecommunications (DIET), University of Rome ``Sapienza'', Via Eudossiana 18, I-00184 Rome, Italy. E-mail: paolo.giannitrapani@uniroma1.it.}, Elio D. Di Claudio\thanks{Elio D. Di Claudio, formerly with Dept. of Information Engineering, Electronics and Telecommunications (DIET), University of Rome ``La Sapienza'', Via Eudossiana 18, I-00184 Rome, Italy.} and Giovanni Jacovitti\thanks{Giovanni Jacovitti, formerly with Dept. of Information Engineering, Electronics and Telecommunications (DIET), University of Rome ``La Sapienza'', Via Eudossiana 18, I-00184 Rome, Italy.}}

%\markboth{IEEE Transactions on Image Processing,~Vol.~xx, No.~xx, xx~xxxx}%
\markboth{PROPOSAL2025}%
{BELE: Blur Equivalent Linearized Estimator}

\maketitle

\begin{abstract}
In the Full-Reference Image Quality Assessment context, Mean Opinion Score values represent subjective evaluations based on retinal perception, while objective metrics assess the reproduced image on the display. Bridging these subjective and objective domains requires parametric mapping functions, which are sensitive to the observer's viewing distance. This paper introduces a novel parametric model that separates perceptual effects due to strong edge degradations from those caused by texture distortions. These effects are quantified using two distinct quality indices. The first is the Blur Equivalent Linearized Estimator, designed to measure blur on strong and isolated edges while accounting for variations in viewing distance. The second is a Complex Peak Signal-to-Noise Ratio, which evaluates distortions affecting texture regions. The first-order effects of the estimator are directly tied to the first index, for which we introduce the concept of \emph{focalization}, interpreted as a linearization term. Starting from a Positional Fisher Information loss model applied to Gaussian blur distortion in natural images, we demonstrate how this model can generalize to linearize all types of distortions. Finally, we validate our theoretical findings by comparing them with several state-of-the-art classical and deep-learning-based full-reference image quality assessment methods on widely used benchmark datasets.
\end{abstract}


\section{Introductive notes}

\IEEEPARstart{I}{mage} Quality Assessment (IQA) can be conducted through subjective or objective methods. Subjective quality is measured using Mean Opinion Score (MOS), which averages ratings assigned by human subjects, while the Difference of Mean Opinion Score (DMOS) quantifies perceived quality loss relative to a reference image.

However, subjective methods are impractical for real-time applications like broadcasting, necessitating objective methods that estimate subjective quality using algorithmic indices. These methods rely on IQA databases, which contain distorted images annotated with empirical MOS/DMOS ratings obtained from human assessments.

Various objective IQA methods have been developed for different protocols \cite{WANG04,SHEIKH06B,CHOW16}. Full Reference (FR) methods require both original and distorted images, Reduced Reference (RR) methods use partial image information, and No Reference (NR) methods rely solely on the degraded image. This study focuses on FR-IQA methods.

IQA methods typically follow a three-stage process. First, local features are extracted based on similarity criteria and theoretical models of visual perception. Next, these features are pooled across the image to compute an IQA metric. Finally, a parametric scoring function maps the metric values onto MOS/DMOS scores, with parameters optimized using empirical data from IQA databases. The VQEG \cite{VQEG00,VQEG03} recommends a logistic function with three to five parameters, while \cite{ITU16B} proposes an S-curve function with three parameters.

The scoring function is influenced by the database images, which are typically natural images captured by optical cameras. Its accuracy depends on experimental settings, particularly viewing distance \cite{LIN11}, as subjective MOS/DMOS scores are based on retinal perception, while objective quality is computed from displayed images. When quality estimates are needed for different viewing distances — such as in auditoria or classrooms — re-training is required for each scenario.

% *** blocco riutilizzabile in revisione ***
%David Marr, in his pioneering work on vision \cite{MARR10}, proposed that the brain processes visual information in stages. It starts by identifying basic features like edges and brightness changes, then combines them into more complex forms, such as surfaces and shapes, considering depth and orientation. Finally, the brain constructs a complete 3D representation of objects or scenes. Marr's model suggests that both local and global distortions can disrupt our visual perception. Our work focuses on the first stage of fine localization.
% ******************************************

% table of notation
\begin{table*}[ht]
	\begin{tabular}{|l|l|}
		\hline
		\multicolumn{2}{|c|}{Main notation}	\\
		\hline \hline
		
		\multicolumn{2}{|c|}{Human Visual System (HVS) model} \\
		\hline
		$\mathbf{p}$ & Point in space, $\mathbf{p}\equiv(p_1;p_2)$.	\\
%		\hline
		$\widetilde{I}\left(\mathbf{p}\right)$, $I\left(\mathbf{p}\right)$ & Luminance images projected onto the retina, representing the reference and degraded images, respectively. \\
%		\hline
		$h\left(\mathbf{p}\right)$ & Complex Laguerre filter as an abstract functional model of the HVS. \\
%		\hline
		$H(\rho,\vartheta)$ & $h\left(\mathbf{p}\right)$ in the frequency domain. Its magnitude is a model of the Contrast Sensitivity Function of the HVS. \\
%		\hline
		$\widetilde{y}\left(\mathbf{p}\right)$, $y\left(\mathbf{p}\right)$ & Visual maps of the reference and degraded images projected onto the retina, respectively. \\
%		\hline
		$\widetilde{\psi}(\mathbf{p})$, $\psi(\mathbf{p})$ & The Positional Fisher Information of the detail for the reference and degraded images, respectively, in the spatial domain. \\
%		\hline
		$\widetilde{\psi}(\rho,\vartheta)$, $\psi(\rho,\vartheta)$ & The Positional Fisher Information of the detail for the reference and degraded images, respectively, in the frequency domain. \\
%		\hline
		${\hat{d}}_{\text{CAN}}(Q;\tau;\xi)$ & The canonical model. \\
%		\hline
		${\hat{d}}_{\text{EMP}}(Q;\tau;\xi)$ & The empirical model, based on the content of the images.	\\
		\hline \hline
		
		\multicolumn{2}{|c|}{System parameters} \\
		\hline
		MOS, DMOS & Subjective quality metric: the Mean Opinion Score (MOS) and the Difference of Mean Opinion Score (DMOS). \\
%		\hline
		Q & Statistical anchor. It defines the concept of worst quality in the experiments. \\
%		\hline
		$\tau$ & The viewing distance. \\
%		\hline
		$\xi$ & The level of Gaussian blur added to the degraded image. \\
%		\hline
		$\zeta \equiv \text{DMOS}$ & The metric required by the conversion function to obtain the equivalent blur values $\xi_{eq}$.	\\
%		\hline
		$\xi_{eq}$ & The equivalent blur, representing any distortion other than Gaussian blur that results in the same DMOS. \\
%		\hline
%		$\Omega_C$ & The set of image points above the natural vision threshold, where strong and isolated edges are analyzed. \\
%		\hline
%		$\Omega_H$ & The set of image points below the natural vision threshold, where textures are analyzed. \\
		\hline \hline
		
		\multicolumn{2}{|c|}{Performance metrics} \\
		\hline
		$\text{BELE}_{\text{cold}}$ & The estimator associated with the quality assessment of strong and isolated edges. \\
%		\hline
		$\text{CPSNR}(\widetilde{y}(\mathbf{p}), y(\mathbf{p}))$ & The estimator associated with the quality assessment of textures. \\
%		\hline  
		$\text{BELE}$ & The final estimator, resulting from the combination of quality assessments for strong and isolated edges and textures. \\
		\hline
	\end{tabular}
	\caption{Main notation for Human Visual System model, system parameters and performance metrics.}
	\label{table:not}
\end{table*}

This work addresses the problem from a theoretical perspective, leveraging a Human Visual System (HVS) model for blur correction in natural images. The model explains orientation selectivity and 2D spatial frequency selectivity using a single complex-valued Virtual Receptive Field (VRF), while explicitly accounting for viewing distance \cite{DICLAUDIO21,GIANNITRAPANI25}.

The analysis is based on the principle that the HVS is optimized for fine pattern localization, given the eye’s macrostructure \cite{DICLAUDIO21}. From an information-theoretic standpoint, the variance of fine localization is an objective measure that depends on blur and can be approximated by the inverse of the Fisher Information regarding pattern position. By leveraging this principle and the spectral properties of natural images, the scoring function for Gaussian blur-induced quality loss can be expressed in a closed form \cite{DICLAUDIO21}, aligning well with empirical DMOS data.

The main contributions of this work include:
\begin{itemize}
	\item{Empirical estimator for strong edges under Gaussian blur: A linearized estimator for blur on strong edges at different viewing distances, extending the approach in \cite{GIANNITRAPANI25} by incorporating image content.}
	\item{Isoluminance filter design: A filtering mechanism applied to reference and blurred image pairs to analyze deviations in subjective quality perception (DMOS) from the canonical estimator.} %Additionally, the filter proves useful in assessing the residual blur after focalization.}
	\item{Linearization through focalization: A novel linearization framework based on Positional Fisher Information (PFI), allowing the model to generalize beyond Gaussian blur to other distortions.}
	\item{Separation of perceptual effects: The introduction of a second index for texture distortions, enabling a more comprehensive quality assessment when combined with the edge-based index.}
\end{itemize}

\begin{figure}[!t]
	\centering
	\includegraphics[width=2.8in]{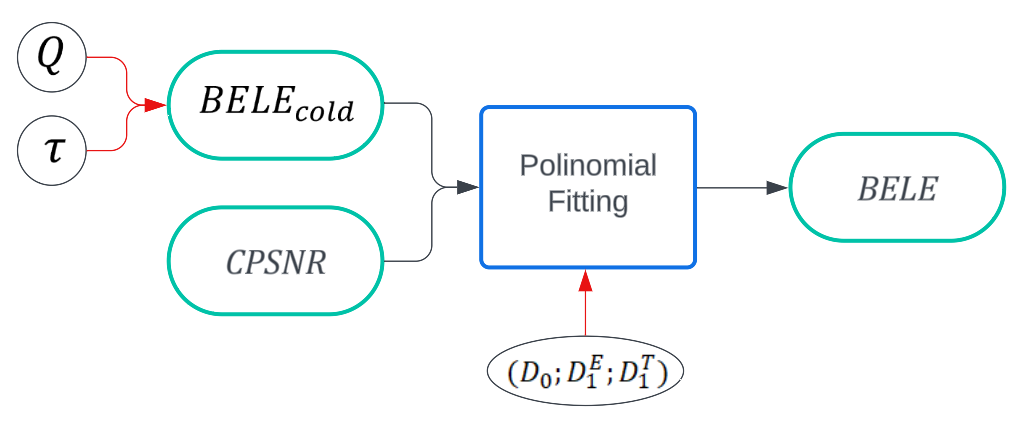}
	\caption{Flowchart illustrating the construction of the BELE estimator. BELE is formed by combining two separately studied indices: $\text{BELE}_{\text{cold}}$ for strong and isolated edges and CPSNR for textures. The process requires five parameters: two ($Q$ and $\tau$) physically modeling blur and three for the polynomial fitting between the indices, representing an affine transformation. This contrasts with the VQEG transformation, which uses a five-parameter logistic function. The affine combination will be detailed in Sec. \ref{sec:Combination of metrics for strong edges and textures}.}
	\label{fig:indices_overall}
\end{figure}

The paper begins with a theoretical background in Sec. \ref{sec:Background}, introducing the concept of the virtual receptive field and the canonical model. Sec. \ref{sec:Empirical estimator of strong edges} presents the empirical estimator for strong edges, demonstrating how direct image content analysis improves accuracy over the canonical estimator, alongside the introduction of the isoluminance filter for visual identification. Sec. \ref{sec:Focusing} extends this approach with the concept of focalization, enabling distortion estimation beyond Gaussian blur and leading to the development of the Blur Equivalent Linearized Estimator (BELE). Sec. \ref{sec:Complex PSNR for textures} addresses second-order effects by introducing a complex PSNR metric for texture analysis. Sec. \ref{sec:Combination of metrics for strong edges and textures} integrates the proposed metrics using affine transformations to handle both strong edges and textures. Sec. \ref{sec:Performance evaluation} evaluates the performance of the proposed methods, while Sec. \ref{sec:Conclusion} summarizes the findings and suggests future research directions. Fig. \ref{fig:indices_overall} provides a flowchart illustrating how BELE is constructed from two indices, one analyzing strong and isolated edges and the other evaluating textures.

\section{Background}
\label{sec:Background}

\noindent We briefly present the \emph{canonical model} for estimating Gaussian blur in natural images, as introduced in \cite{DICLAUDIO21,GIANNITRAPANI25}.% Fig. \ref{fig:canonical_model} illustrates the computational pipeline, detailing the operational layers leading to the canonical model. 
Tab. \ref{table:not} summarizes the main notations used.  

The Human Visual System (HVS) processes visual information through \emph{receptive fields} \cite{HUBEL95}, abstractly modeled as the convolution of the luminance image $I(\mathbf{p})$ with the Virtual Receptive Field (VRF) $h(\mathbf{p})$:
\begin{equation}
	y(\mathbf{p}) = I(\mathbf{p}) \ast h(\mathbf{p}).
	\label{eqn:visualmap}
\end{equation}

The output $y\left(\mathbf{p}\right)$ will be referred to as the \emph{visual map}. According to \cite{DAUGMAN83}, $h(\mathbf{p})$ is the complex filter accounting for the orientation and frequency selectivity of the HVS.% It is expressed as:
%\begin{equation}
%	h(\mathbf{p}) = Re{\left\{h(\mathbf{p})\right\}} + jIm{\left\{h(\mathbf{p})\right\}}.
%\end{equation}

%We show in Fig. \ref{fig12} the complex Laguerre filters separated into their real and imaginary parts, in polar coordinates \cite{DICLAUDIO21}:
%The complex Laguerre filters, in polar coordinates \cite{DICLAUDIO21} is:
%\begin{equation}
%	h(r,\varphi) = \frac{r}{2\pi s_G^2}e^{-\frac{r^2}{2{s_G}^2}}e^{j\varphi}.
%	\label{eqn:hmodel}
%\end{equation}

%Its Fourier transform, the Virtual Neural Transfer Function, is:
The Fourier transform of the complex Laguerre filters, in polar coordinates, is the Virtual Neural Transfer Function \cite{DICLAUDIO21}:
\begin{equation}
	H(\rho,\vartheta) = j2\pi\rho e^{j\vartheta}e^{-s_G^2\rho^2}.
\end{equation}

%This function models the Contrast Sensitivity Function (CSF) of the HVS \cite{MANNOS74}, shown in Fig. \ref{fig14}.
This function models the Contrast Sensitivity Function (CSF) of the HVS \cite{MANNOS74}. The spread $s_G = 2.5$ arcmin aligns with experimental data \cite{CAMPBELL65}, peaking at 8.5 cycles/degree and dropping $\sim$40 dB at 30 cycles/degree.  

Fisher Information (FI) \cite{TREES92} quantifies estimation precision of the image projected onto the retina. The Positional Fisher Information (PFI) \cite{NERI04} for a pristine image is:
\begin{equation}
	\widetilde{\psi}(\mathbf{p}) = \frac{\widetilde{\lambda}(\mathbf{p})}{\sigma_V^2},
\end{equation}
where $\widetilde{\lambda}(\mathbf{p})$ is the smoothed gradient energy of the detail:
\begin{equation}
	\widetilde{\lambda}(\mathbf{p}) = \sum_{\mathbf{q}}{w_\mathbf{p}(\mathbf{q})^2\left|\widetilde{y}\left(\mathbf{p}-\mathbf{q}\right)\right|^2}.
	\label{eqn:lamda_tilde}
\end{equation}
and $\sigma_V^2$ is the additive Gaussian noise variance. $w_\mathbf{p}(\mathbf{q})$ is a sampling window centered on $\mathbf{p}$.

Using Parseval’s theorem, in the frequency domain, the expected value of the PFI over natural images is \cite{DICLAUDIO21}:
\begin{equation}
	\widetilde{\mathrm{\Psi}} = \frac{1}{\sigma_V^2}\int_{0}^{2\pi}{f\left(\vartheta\right)\int_{0}^{+\infty}{\left|G(\rho,\vartheta)\right|^2\rho d\rho d\vartheta}} \; ,
\end{equation}
where $G(\rho,\theta)=e^{-s_G^2\rho^2}$.

For blurred images, considering the Optical Transfer Function (OTF) $B(\rho,\vartheta)$:
\begin{equation}
	\mathrm{\Psi} = \frac{1}{\sigma_V^2}\int_{0}^{2\pi}{f\left(\vartheta\right)\int_{0}^{+\infty}{\left|G(\rho,\vartheta)\right|^2\left|B(\rho,\vartheta)\right|^2\rho d\rho d\vartheta}}.
\end{equation}

In the case of isotropic Gaussian blur, $B(\rho,\vartheta) = e^{-s_B^2\rho^2}$, leading to:
\begin{equation}
	\frac{\mathrm{\Psi}}{\widetilde{\mathrm{\Psi}}} = \frac{s_G^2}{s_G^2 + s_B^2}.
\end{equation}

The increase in positional uncertainty follows Weber’s law \cite{DICLAUDIO21} and is given by:
\begin{equation}
	\varepsilon\left(\xi\right) = 1 - \sqrt{\frac{1}{1+\xi^2}},
\end{equation}
where $\xi = s_B / s_G$ is the \emph{normalized blur}.

The quality loss is mapped onto the DMOS scale, and the model is completed by introducing two parameters: $Q$, defining the worst quality, and $\tau$, representing viewing distance, leading to the final form of the \emph{canonical} DMOS estimator \cite{GIANNITRAPANI25}:
\begin{equation}
	{\hat{d}}_{\text{CAN}}(Q;\tau;\xi) = 100 \times Q \times \left(1 - \frac{1}{\sqrt{1+\displaystyle\frac{\xi^2}{\tau^4}}}\right).
	\label{eqn:dblur}
\end{equation}

Popular databases provide direct or indirect viewing distance information. Using empirical DMOS data and known $\xi$, $\tau$ and $Q$ are estimated via regression of \eqref{eqn:dblur}.

\section{Empirical estimator of strong edges}
\label{sec:Empirical estimator of strong edges}

\begin{figure}[!t]
	\centering
	\includegraphics[width=1.65in]{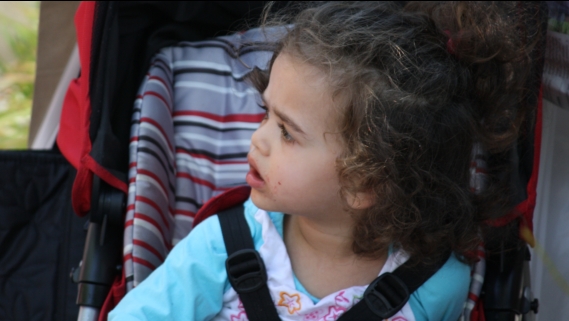}
	\includegraphics[width=1.65in]{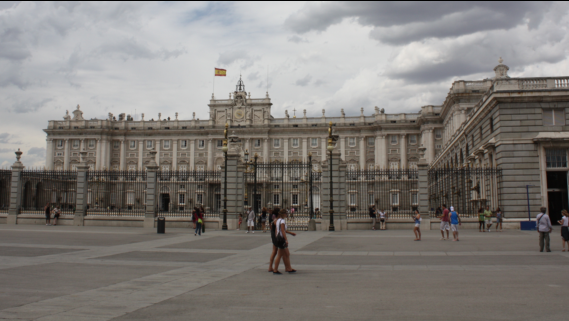} \vspace{1mm} \\
	\includegraphics[width=1.65in]{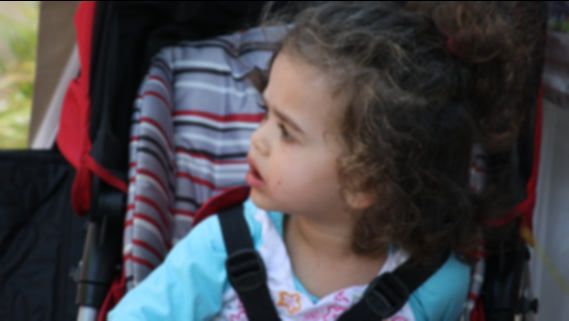}
	\includegraphics[width=1.65in]{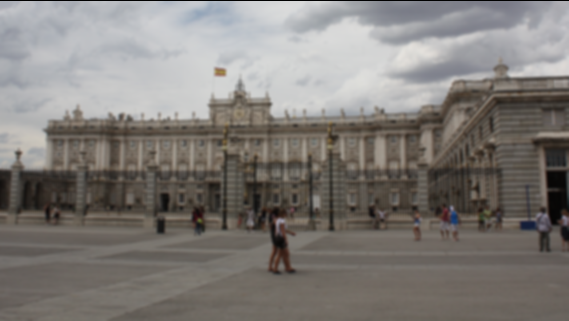} \vspace{1mm} \\
	\includegraphics[width=1.65in]{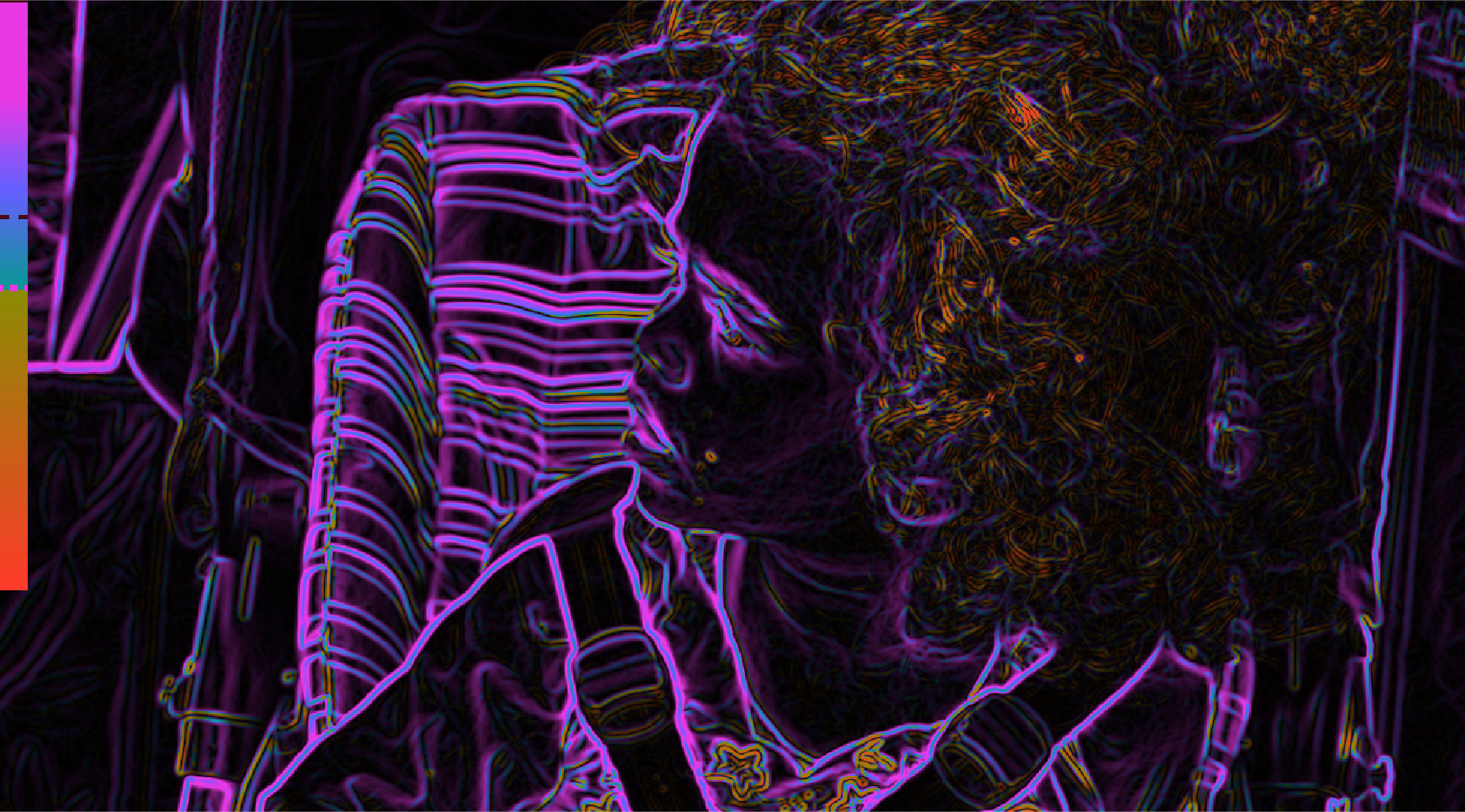}
	\includegraphics[width=1.65in]{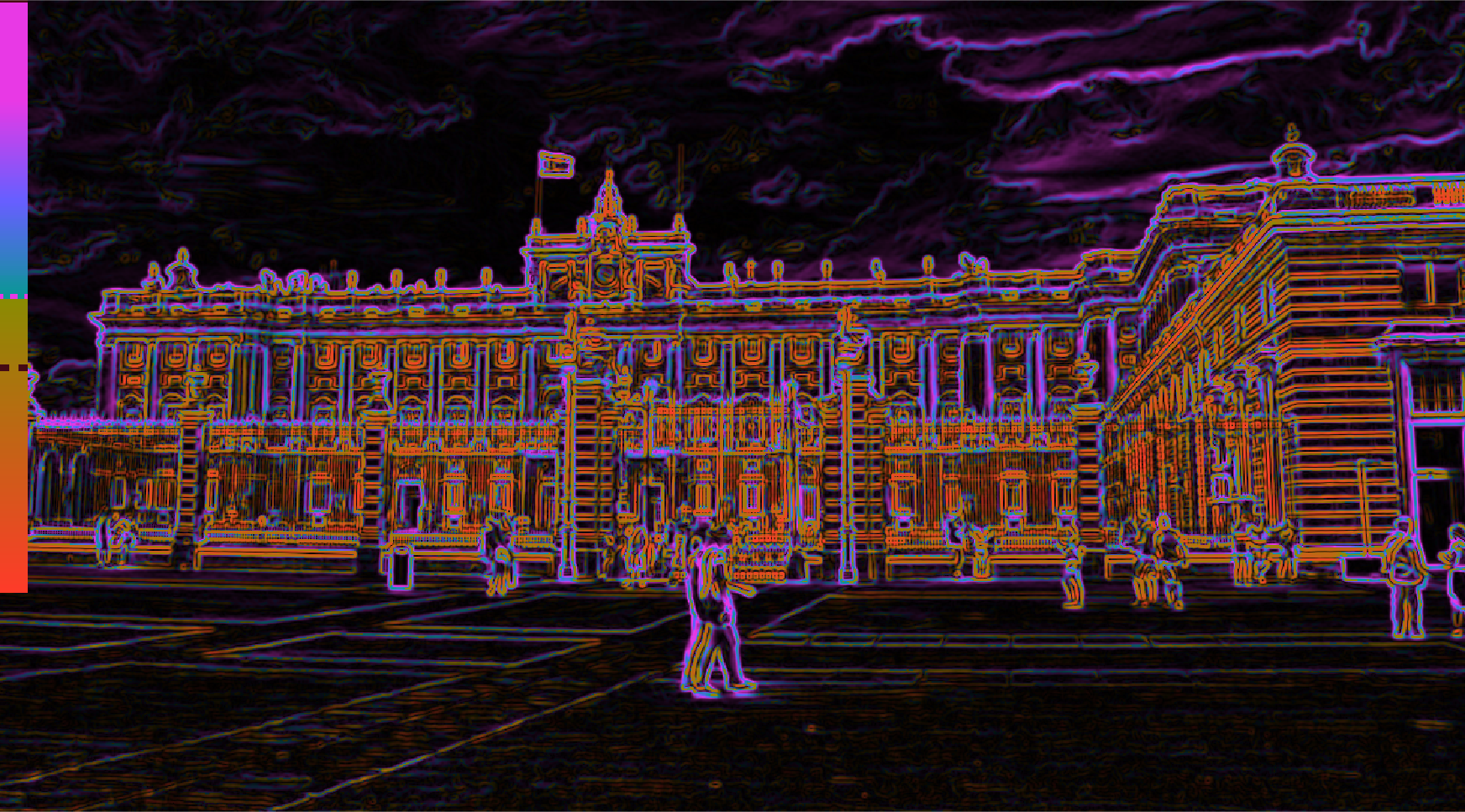} \vspace{1mm} \\
	\caption{Upper row: two original images, "Babygirl" and "Palace2". Second row: Gaussian blurred images with $s_B=2.4$. Lower row: certainty maps. Purple dominance (left column) indicates higher certainty in high-quality images, while red dominance (right column) signifies greater loss in lower-quality images. The isoluminance colors maintain a constant intensity at the same edge level, so, only the weight defines the local contrast.}
	\label{fig:babygirl_palace2}
\end{figure}

\begin{figure}[!t]
	\centering
	\includegraphics[width=1.65in]{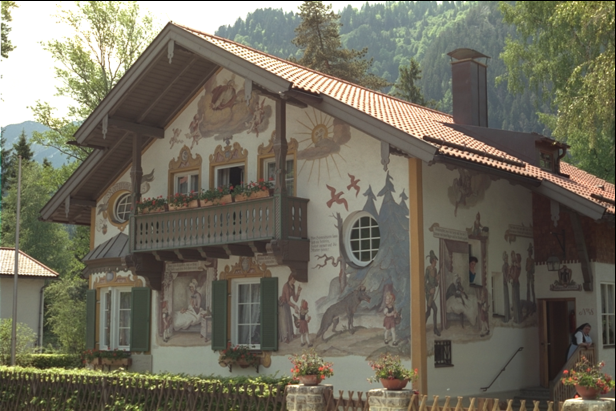}
	\includegraphics[width=1.65in]{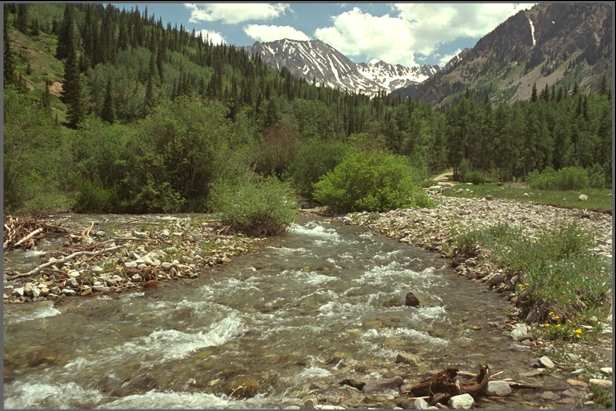} \vspace{1mm} \\
	\includegraphics[width=1.65in]{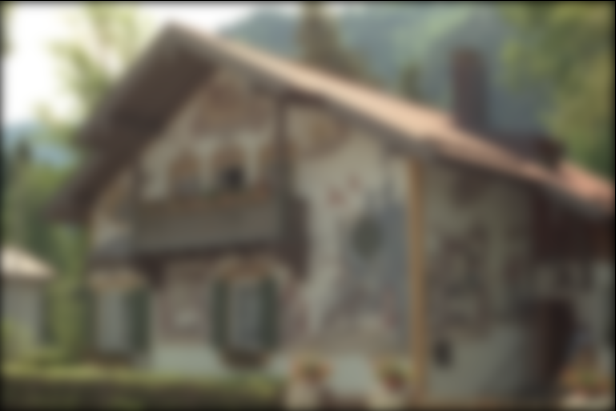}
	\includegraphics[width=1.65in]{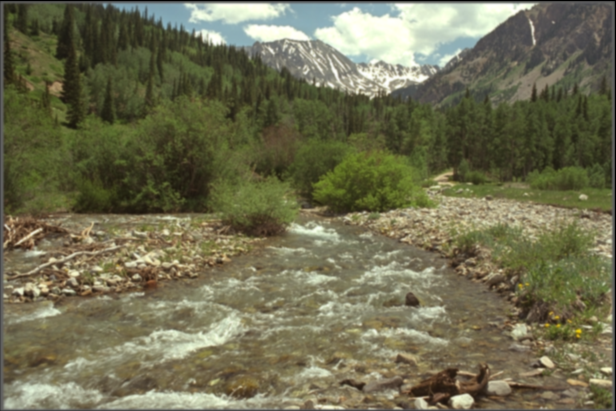} \vspace{1mm} \\
	\includegraphics[width=1.65in]{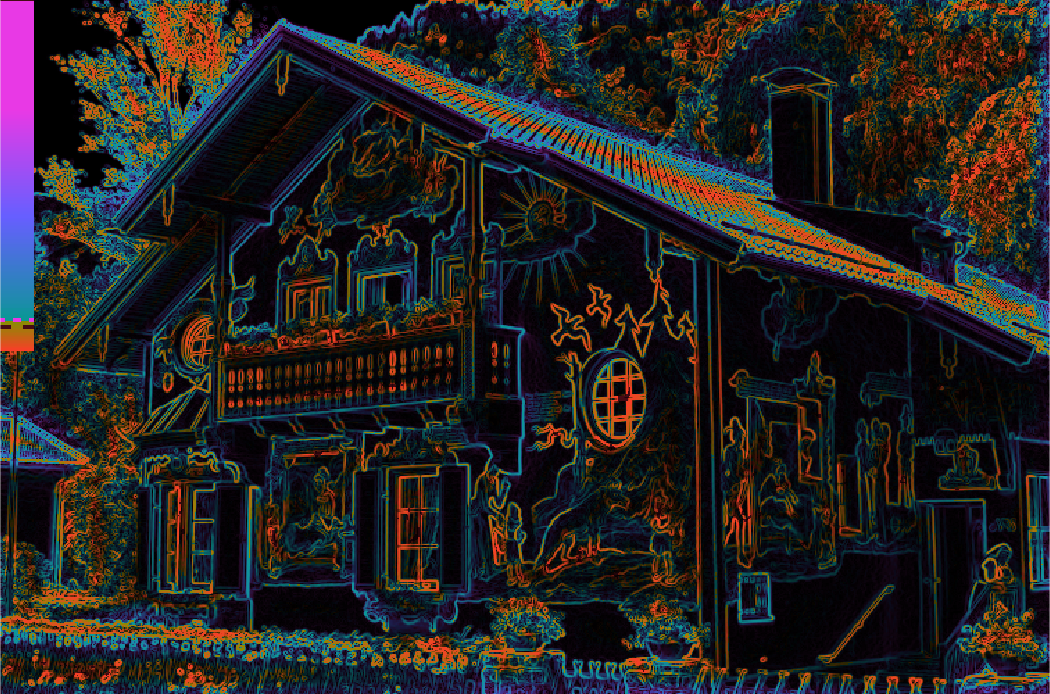}
	\includegraphics[width=1.65in]{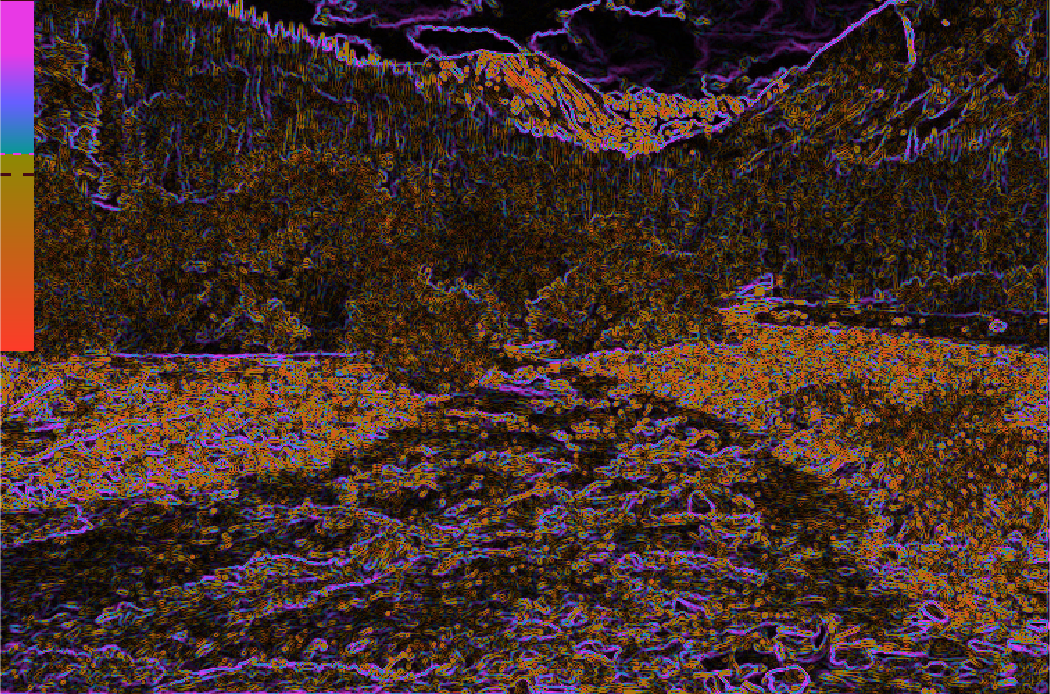} \vspace{1mm} \\
	\caption{Upper row: two original images, "Paintedhouse" and "Stream". Second row: Gaussian blurred images with $s_B=7.7$ and $s_B=1.0$. Lower row: certainty maps. No dominance of purple or red is observed. In the first image, "Paintedhouse," a widespread cyan color suggests that subjective perception closely aligns with the canonical estimator.}
	\label{fig:paintedhouse_stream}
\end{figure}

\noindent Building on the canonical model (\ref{eqn:dblur}), the empirical estimator incorporates variations in subjective quality due to image content.% Fig. \ref{fig:empirical_model} outlines the computational pipeline, detailing the operational layers leading to the empirical model.

The empirical DMOS estimator can be defined as:
\begin{equation}
	{\hat{d}}_{\text{EMP}}(Q;\tau;\xi) = 100\times Q \times \left(1 - \sqrt{R}\right),
	\label{eqn:d_emp}
\end{equation}
where $Q$ is the gain parameter that scales the scoring system according to the considered database, and
\begin{equation}
	R = \frac{1}{N_{\Omega_C}} \sum_{\textbf{p} \in \Omega_C} \frac{\lambda(\textbf{p})}{\tilde{\lambda}(\textbf{p})}.
\end{equation}
%with
%\begin{IEEEeqnarray}{l}
%	\lambda(\textbf{p}) = \sum_{\textbf{q}} w_{\textbf{p}}(\textbf{q})^2 |y(\textbf{p} - \textbf{q})|^2, \\
%	\tilde{\lambda}(\textbf{p}) = \sum_{\textbf{q}} w_{\textbf{p}}(\textbf{q})^2 |\tilde{y}(\textbf{p} - \textbf{q})|^2.
%\end{IEEEeqnarray}
%where $y(\textbf{p})$ is the visual map of the blurred image, and $\tilde{y}(\textbf{p})$ represents the corresponding reference image.
with $\lambda(\textbf{p}) = \sum_{\textbf{q}} w_{\textbf{p}}(\textbf{q})^2 |y(\textbf{p} - \textbf{q})|^2$ and $\tilde{\lambda}(\textbf{p})$ is defined in (\ref{eqn:lamda_tilde}). $y(\textbf{p})$ is the visual map of the blurred image, and $\tilde{y}(\textbf{p})$ represents the corresponding reference image.

Let us define the \emph{certainty map} $M(\textbf{p})$ as:
%\begin{equation}
%	M(\textbf{p}) = 
%	\begin{cases} 
%		\frac{|y(\textbf{p})|}{|\tilde{y}(\textbf{p})|} & \text{if} \quad \frac{|y(\textbf{p})|}{|\tilde{y}(\textbf{p})|} \leq 1, \\
%		0 & \text{elsewhere}.
%	\end{cases}
%\end{equation}
\begin{equation}
	M(\textbf{p}) = \frac{|y(\textbf{p})|}{|\tilde{y}(\textbf{p})|},
	\label{eqn:M_thre}
\end{equation}
where $M(\textbf{p})$ quantifies the certainty of the visual information at the pixel location $\textbf{p}$.

%The empirical linear estimator is obtained by considering the portion of the image where natural and strong edges dominate. Specifically, $\Omega_C$ is the set of points where $M(\textbf{p}) \geq \overline{M}$, with
%\begin{equation}
%	\overline{M} = \sqrt{\frac{1}{1 + \xi^2}}.
%\end{equation}
%This condition ensures that the estimator takes into account the most significant visual features, those corresponding to strong edges.

\begin{definition}  
	The \textbf{certainty region} is defined as the subset of the image where strong and isolated edges dominate, ensuring reliable estimation. Formally, this region is given by the set:
	\[
	\Omega_C = \{ \textbf{p} \in \Omega \ | \ M(\textbf{p}) \geq \overline{M} \},
	\]
	where the threshold \(\overline{M}\) is defined as:
	\[
	\overline{M} = \sqrt{\frac{1}{1 + \xi^2}}
	\]
	and $C$ stands for cold area.
\end{definition}

%As shown in Fig. \ref{fig:canonical_model}, the threshold $\overline{M}$ is directly derived from the canonical model.
The threshold $\overline{M}$ is directly derived from the canonical model.

% distanza di visione nello stimatore empirico
To account for viewing distance variations, the empirical estimator scales the convolutional operator used for feature extraction. The kernel size is adjusted by the parameter $\gamma = 1/\tau$, affecting the Gaussian kernel's standard deviation as $\sigma = \sigma_{\text{initial}} / \gamma^2$. This ensures that gradient-based feature extraction reflects perceptual effects at different distances, dynamically adapting to spatial frequency changes and maintaining consistent subjective quality predictions.

% filtro ottico a isoluminanza

% *** blocco riutilizzabile in revisione ***
%To analyze the image content and understand the impact of blur on the edges, we refer to the isoluminance filter used in the calibration case for isolated edges.
% ******************************************

To analyze image content and the impact of blur on edges, we use the isoluminance filter based on the certainty map $M(\textbf{p})$ from (\ref{eqn:M_thre}). Cyan highlights isolated, strong edges at the natural vision threshold where $M(\textbf{p}) = \overline{M}$. As certainty increases ($M(\textbf{p}) > \overline{M}$), values shift toward purple, marking points in $\Omega_C$. Conversely, weaker edges and overlapping regions appear in red, indicating blur-dominated areas ($M(\textbf{p}) < \overline{M}$), forming the hot area set $\Omega_H$.

In the isoluminance filter images in Fig. \ref{fig:babygirl_palace2} and Fig. \ref{fig:paintedhouse_stream}, the colorbar on the left includes two key markers. The purple dotted line represents the theoretical index from the canonical model (Sec. \ref{sec:Background}), distinguishing the natural loss threshold based on the blur level $s_B$ at a fixed viewing distance. This index is always linked to the reference color cyan. The dark dotted line, instead, corresponds to the DMOS values, reflecting subjective image judgment. Fig. \ref{fig:babygirl_palace2}, from the LIVE MD dataset \cite{JAYARAMAN12}, illustrates the highest error cases at both ends of the LIVE MD quality scale, while Fig. \ref{fig:paintedhouse_stream}, from the LIVE DBR2 dataset, showcases two cases of effective correction.

In the left column of Fig. \ref{fig:babygirl_palace2}, the subjective judgment exceeds the theoretical index, indicating a more favorable rating than the objective quality, as shown by the dominance of purple. Conversely, in the right column, the subjective judgment is significantly worse than the canonical estimate, with the dominance of red visually highlighting the perceived degradation. In Fig. \ref{fig:paintedhouse_stream}, neither purple nor red dominates. In the first image, "Paintedhouse," the prevalent cyan hue suggests that subjective perception aligns closely with the canonical estimator.

Tab. \ref{table:indexes} presents the initial values obtained using only the canonical estimator ${\hat{d}}_{\text{CAN}}(Q;\tau;\xi)$ and the corrected values provided by the empirical estimator ${\hat{d}}_{\text{EMP}}(Q;\tau;\xi)$. Here, the empirical estimator demonstrates its ability to align closely with the DMOS values, even for high levels of blur, outperforming the canonical estimator in accuracy.

% *** blocco riutilizzabile in revisione ***
%Theoretical derivations regarding the PFI for strong and isolated edges can be further explored in Appendix \ref{app:Theoretical considerations for isolated edges}.
% ******************************************

% theoretic/dmos/empric indexes - table
% per la verifica di questi indici lanciare la funzione imgpartitionblur.m o imgpartitionblur_concise.m e vedere i vettori "refimgax_ord", "sBvec_ord", "SBDItheoretic_ord", "SBDIactual_corr_ord" e "dmosnew_ord".
\begin{table}[!t]
	\centering
	\caption{Initial and corrected values: canonical estimator ${\hat{d}}_{\text{CAN}}(Q;\tau;\xi)$, empirical estimator ${\hat{d}}_{\text{EMP}}(Q;\tau;\xi)$, and DMOS values. The empirical estimator demonstrates superior alignment with DMOS, even for high levels of blur, surpassing the accuracy of the canonical estimator.}
	\label{table:indexes}
	\scalebox{0.9}{
		\begin{tabular}{llcccc}
			\noalign{\smallskip}
			\toprule
			Dataset & Image name & $\text{$s_B$}$ & ${\hat{d}}_{\text{CAN}}(Q;\tau;\xi)$ & ${\hat{d}}_{\text{EMP}}(Q;\tau;\xi)$ & DMOS \\
			\midrule
			LIVE MD & Babygirl & $2.4$ & $52.21$ & $34.70$ & $40.00$ \\
			LIVE MD & Palace2 & $2.4$ & $53.42$ & $57.76$ & $66.37$ \\
			\midrule
			DBR2 & Paintedhouse & $7.7$ & $83.27$ & $86.29$ & $86.27$ \\
			DBR2 & Stream & $1.0$ & $38.72$ & $46.99$ & $46.66$ \\
			\bottomrule
		\end{tabular}
	}
\end{table}

\section{Focusing}
\label{sec:Focusing}

\noindent So far, we have addressed Gaussian blur estimation, incorporating both the canonical model and image content. We have observed deviations in subjective perception from the theoretical values and identified the set $\Omega_C$ as crucial for assessing the degradation of strong, isolated edges.  

Now, we extend the focus to estimating distortions beyond Gaussian blur. Specifically, the task of focusing — realigning all distortions using the canonical Gaussian blur model — fully replaces the VQEG rectification step found in previous models.

%Fig. \ref{fig:empirical_model} illustrates the computational pipeline, detailing the operational layers that lead to the focusing model. It also highlights the integration of these models with the sensitivity-based power computation to derive the final estimator.

Full-Reference IQA methods typically employ a scoring function, $\hat{m}(\zeta)$, to map the objective IQA metric $\zeta$ to an estimated MOS/DMOS value. This function is parametric, accounting for threshold and saturation effects, with parameters optimized via non-linear regression on empirical data. While additional parameters enhance dataset-specific accuracy, they may reduce generalization to new datasets.

A widely used scoring function is the VQEG model, $\hat{m}_{VQEG}(\zeta)$, which applies a logistic function \cite{VQEG00, VQEG03}:  
\begin{equation}
	\hat{m}_{VQEG}(\zeta) = \beta_1 \left[\frac{1}{2} - \frac{1}{1 + e^{\beta_2 (\zeta - \beta_3)}}\right] + \beta_4 \zeta + \beta_5,
\end{equation}  
where $\zeta$ is the IQA metric, $\hat{m}(\zeta)$ the estimated DMOS, and $\beta_i$ are parameters optimized by minimizing the Euclidean distance between empirical and estimated DMOS values. A simpler three-parameter S-shaped function has also been introduced in \cite{ITU16B}.

\begin{figure}[!t]
	\centering
	\includegraphics[width=2.6in]{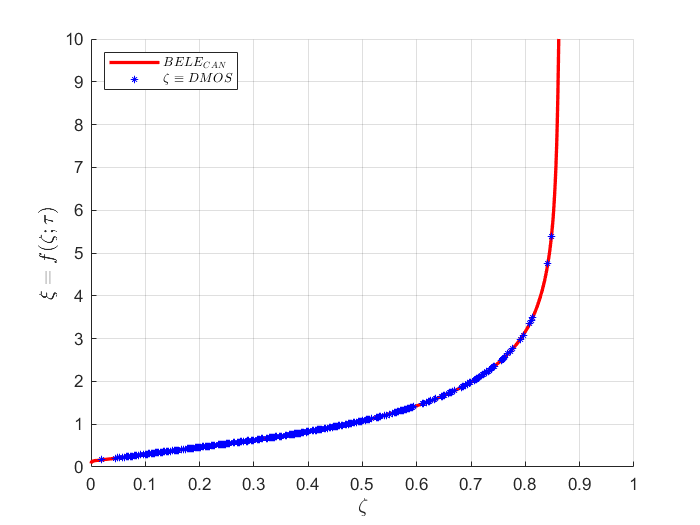}
	\caption{The conversion function of $\text{BELE}(\zeta)$ (red curve), along with the equivalent blur values obtained by mapping the DMOS values of all distortions in the LIVE DBR2 dataset onto the conversion function (blue stars).}
	\label{fig:conversion_function_BELE}
\end{figure}

%\subsection{Focused BELE for cold areas}
%\label{subsec:Focused BELE for cold areas}

In Shannon's information theory, noise refers to any factor that degrades the quality of transmitted information \cite{cover2006elements}. Similarly, in IQA, distortions like Gaussian blur can be seen as noise that corrupts the reference image, reducing its quality. These distortions impair the perception of the original image, much like how noise affects signal fidelity in communication.

For example, the VIF method \cite{SHEIKH06} computes the cumulative contribution of terms that quantify Shannon information losses. In this context, increased noise is treated similarly to the energy reduction caused by blurring, both leading to information degradation.

In \cite{GIANNITRAPANI25}, we showed that any IQA metric can be decomposed into a metric conversion rule and the canonical IQA method. This means that the conversion function $\xi(\zeta; \tau)$ is valid even when metric values $\zeta$ are affected by various factors, not just Gaussian blur. These factors produce equivalent results for $\hat{d}_{\text{CAN}}(\xi) = \hat{d}_{\text{IQA}}(\zeta)$, highlighting the broad applicability of the conversion method.

The conversion function maps each IQA estimate to an equivalent blur value $\xi_{eq} = f(\zeta; \tau)$, linking the metric values to their blur interpretation. In the empirical estimator, this introduces the concept of focusing with an equivalent model, analyzing the effects of the distortion and the equivalent blur separately. The equivalent model applies Gaussian blur to the reference image, matching the equivalent blur level, which simplifies the subsequent distortion estimation.

The blur level for the focusing index is calculated using the inverse of the canonical model in terms of DMOS:
\begin{equation}
	\xi_{eq} = \tau^2 \cdot \sqrt{\frac{1}{\left(1 - \frac{\text{DMOS}}{100 \times Q}\right)^2} - 1}
	\label{eqn:xi_inverse}
\end{equation}

Fig. \ref{fig:conversion_function_BELE} illustrates the conversion function ${\hat{d}}_{\text{EMP}}(\zeta) = {\hat{d}}_{\text{CAN}}(\xi)$ (red curve, see \cite{GIANNITRAPANI25}), alongside the equivalent blur values $\xi_{eq}$ (blue stars), obtained by mapping the DMOS values of distortions in the LIVE DBR2 dataset \cite{SHEIKH06B}. This comparison confirms the alignment between theoretical predictions and empirical data via the concept of equivalent blur, as further examined in the sensitivity analysis.% Fig. \ref{fig:canonical_model} shows the use of the conversion function to derive $\xi_{eq}$ from $\zeta$.

%\begin{figure}[!t]
%	\centering
%	\includegraphics[width=3.6in]{DBR2_blur_sensitivity.png}
%	\caption{Sensitivity study to find lambda powers at minimum RMSE. The x-axis shows the pairs of lambda powers used for the distortion index for the LIVE DBR2 dataset, which contains images with Gaussian blur, and the focusing index with equivalent blur, as shown in (\ref{eqn:distortion_index}) and (\ref{eqn:blureq_index}), respectively. The red point indicates the pair with the minimum RMSE, where the Gaussian blur power is $\text{blur}_{pow} = 0.65$ and the equivalent blur power is $\text{blur-equivalent}_{pow} = 1.35$.}
%	\label{fig:sensitivity_DBR2}
%\end{figure}

The Blur Equivalent Linearized Estimator (BELE) for cold area (the set $\Omega_C$ of strong and isolated edges) is given by:
\begin{equation}
	\text{BELE}_{\text{cold}} = 100\times Q \times \left[ 1 - \left(1 - \hat{d}_{\text{distortion}}\right)\left(1 - \hat{d}_{\xi_{eq}}\right) \right]
	\label{eqn:bele_cold}
\end{equation}
where:
\begin{equation}
	\hat{d}_{\text{distortion}} = 1 - \sqrt{\frac{1}{N_{\Omega_C}} \sum_{\textbf{p} \in \Omega_C} \left(\frac{\lambda(\textbf{p})}{\tilde{\lambda}(\textbf{p})}\right)^{0.65}}
	\label{eqn:distortion_index}
\end{equation}
is the generic distortion estimation term,
\begin{equation}
	\hat{d}_{\xi_{eq}} = 1 - \sqrt{\frac{1}{N_{\Omega_C}} \sum_{\textbf{p} \in \Omega_C} \left(\frac{\lambda_w(\textbf{p})}{\tilde{\lambda}(\textbf{p})}\right)^{1.35}}
	\label{eqn:blureq_index}
\end{equation}
is the focusing term, where $\lambda_w(\textbf{p}) = \sum_\textbf{q} w_\textbf{p}(\textbf{q})^2 |y_{w \delta}(\textbf{p} - \textbf{q})|^2$ and $y_{w \delta}$ is the visual map for the image with the equivalent blur, calculated at the actual distance.

\begin{figure}[!t]
	\centering
	\includegraphics[width=2.9in]{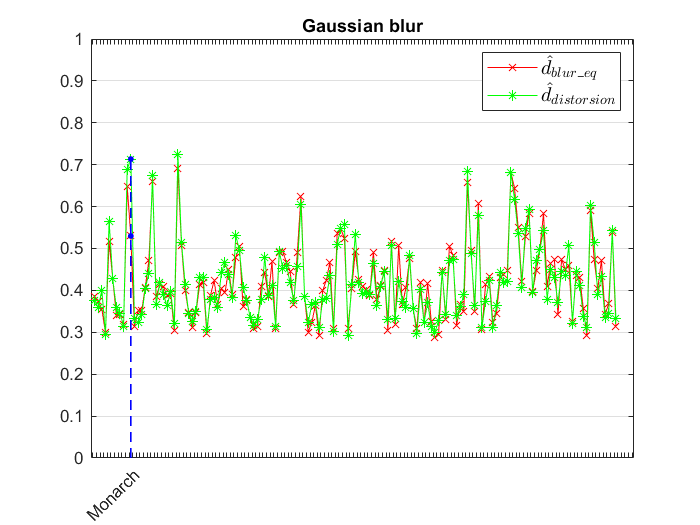}
	\caption{The variation in distortion levels between images affected by Gaussian blur in the LIVE DBR2 dataset (green markers) and their corresponding images with equivalent blur (red markers). This quantifies how the distortion in the equivalent blur image deviates from the expected Gaussian blur distortion in the LIVE DBR2 dataset. The worst-case scenario is highlighted by the “Monarch” image with $s_B = 11.3$, where $\hat{d}_{\xi_{eq}}$ and $\hat{d}_{\text{distortion}}$ are marked by two blue points, showing an approximately 18\% deviation in the prediction.}
	\label{fig:blur_eq-distorsion_DBR2}
\end{figure}

%\begin{figure}[!t]
%	\centering
%	\includegraphics[width=2.8in]{DBR2_blureq-distorsion_scatter.png}
%	\caption{Scatter plot showing the alignment between $\hat{d}_{\xi_{eq}}$ and $\hat{d}_{\text{distortion}}$ for the Gaussian blur from the LIVE DBR2 dataset. The training set includes 145 images, achieving $\text{RMSE}=0.0289$, $\text{SROCC}=0.95486$, and $\text{LCC}=0.95138$. The worst-case “Monarch” with $s_B=11.3$ is represented by the point with the largest deviation in the scatter plot.}
%	\label{fig:blur_eq-distorsion_DBR2_scatter}
%\end{figure}

The estimator integrates both the direct effects of distortion (true degradation) and the compensatory effects of focusing (modeled via equivalent blur). The powers of $\lambda$ in the distortion and focusing terms were optimized through a sensitivity study, aligning Gaussian blur distortions in test images with their equivalent blur added in the reference. This fine-tuning minimizes the Root Mean Square Error (RMSE), ensuring that the empirical model in (\ref{eqn:d_emp}) effectively replaces the Gaussian blur in the LIVE DBR2 dataset, aligning images with the canonical model at the same blur level $s_B$. This approach ensures consistency with the canonical model and empirically validates the concept of equivalent blur \cite{GIANNITRAPANI25}.

The LIVE DBR2 dataset, containing diverse blur levels, was used to validate the equivalence between the index with equivalent blur $\hat{d}_{\xi_{eq}}$ and the distortion index $\hat{d}_{\text{distortion}}$ for Gaussian blur. This process, interpreted as focusing, assigns a weight to the estimation, effectively aligning it with the canonical model.

%Fig. \ref{fig:sensitivity_DBR2} shows the sensitivity analysis optimizing $\lambda$ powers to minimize RMSE. 
Fig. \ref{fig:blur_eq-distorsion_DBR2} compares distortion levels in LIVE DBR2 (green markers) with their equivalent blur counterparts (red markers), highlighting differences from expected Gaussian blur distortion. The metrics demonstrate a strong alignment between $\hat{d}_{\xi_{eq}}$ and $\hat{d}_{\text{distortion}}$, with an RMSE of 0.0289, Spearman Rank Order Correlation Coefficient (SROCC) of 0.95486, and Pearson Linear Correlation Coefficient (PLCC) of 0.95138. These results confirm the effectiveness of the equivalent blur model in improving distortion estimation accuracy.

It is important to emphasize that the level of Gaussian blur $\xi_{eq}$, used to calculate the focusing index is significantly lower than the Gaussian blur present in the test image, specifically by a factor of $s_G = 2.5$. This relationship arises from the resolution chosen for the HVS model, where $\xi = \frac{s_B}{s_G}$.

The factor of $s_G = 2.5$ reflects a scaling between the Gaussian blur applied to the test image ($s_B$) and the equivalent blur used for focusing ($\xi_{eq}$). This explains why there is a higher power in the focusing term.

\section{Complex PSNR for textures}
\label{sec:Complex PSNR for textures}

\noindent The study of \emph{textures} focuses on weak, closely spaced, multi-directional edges that remain above noise but are not fully resolved by the HVS. These \emph{visible textures}, clusters of closely packed edges, differ from isolated edges in key aspects.  

First, the PFI distribution in textured regions is more uniform, as edges are densely packed in multiple directions, preventing separation. In contrast, strong isolated edges concentrate PFI in specific locations. Second, subjective perception of textures is inherently complex—observers struggle to assess blur due to high gradient density and directional ambiguity \cite{LANDY04}. As edge spacing decreases, PFI increases, heightening perceptual uncertainty.  

These factors significantly influence DMOS variability. In texture-dominated regions, DMOS values fluctuate widely, reflecting the challenge human observers face in consistently judging image quality.

To analyze textures, we introduce the Complex Peak Signal-to-Noise Ratio (CPSNR), which quantifies differences in visual maps based on gradient energy distributions. This metric captures fine variations in texture structures.  

The analysis is restricted to perceptually significant details that remain unresolved by the HVS, focusing exclusively on the set $\Omega_H$ (Sec. \ref{sec:Empirical estimator of strong edges}), which includes points below the natural vision threshold.

The $\text{CPSNR}(\widetilde{y}(\mathbf{p}), y(\mathbf{p}))$ is obtained by computing the normalized Mean Squared Error (MSE) directly from the complex visual maps as
\begin{equation}  
	\text{MSE} = \frac{\sum_{i,j} |\widetilde{y}(\mathbf{p})(i,j) - y(\mathbf{p})(i,j)|^2}{\max\limits_{\mathbf{p} \in \Omega_H} (|\widetilde{y}(\mathbf{p})|^2, |y(\mathbf{p})|^2)}.  
\end{equation}

\section{Combination of metrics for strong edges and textures}
\label{sec:Combination of metrics for strong edges and textures}

\noindent The CPSNR metric influences overall quality estimation as a second-order component relative to the empirical estimator described in Sec. \ref{sec:Empirical estimator of strong edges}. This aligns with David Marr’s hierarchical model of visual processing \cite{MARR10}, which emphasizes the dominance of primary components in perception.

Marr's framework highlights the importance of primary edges and contours in shaping visual scene representation. These features drive the empirical estimator's performance by contributing most significantly to the PFI. Meanwhile, CPSNR captures fine variations in texture and small-scale details, serving as a secondary but essential analytical layer.

Integrating CPSNR into quality estimation frameworks follows Marr’s principle that perception arises from primary and secondary components. The empirical estimator corresponds to the "primal sketch," which encodes dominant contours, while CPSNR refines the analysis by addressing textures and finer gradients.

The integration is performed through polynomial fitting, where a polynomial function combines the two indices into a single quality score. The fitting process optimizes polynomial weights to minimize the error between predicted and subjective DMOS scores, using a robust least-squares approach to account for perceptual relevance.

This method assumes that strong edges, represented by $E \equiv \text{BELE}_{\text{cold}}$ in (\ref{eqn:bele_cold}), and textures, represented by $T \equiv \text{CPSNR}(\widetilde{y}(\mathbf{p}), y(\mathbf{p}))$, are perceptually distinct. Their separation into regions $\Omega_C$ and $\Omega_H$ aligns with the natural vision threshold, ensuring specialized quality assessment for each subset of points.

Mathematically, this distinction can be expressed through a Taylor series expansion of the quality function $f(E, T; \mathbf{D})$ around any reference point $(E_0, T_0)$:
\begin{equation}
	\frac{\partial^{m+n} f(E, T; \mathbf{D})}{\partial E^m \partial T^n} \approx 0, \quad \text{for } m, n \geq 1,
\end{equation}
which implies negligible cross-sensitivity between the two indices. $\mathbf{D}$ is the vector of the adaptation coefficient represented by the DMOS. Consequently, $f(E, T; \mathbf{D})$ can be expressed in a decoupled form:
\begin{equation}
	f(E, T; \mathbf{D}) = f_E(E; \mathbf{D}_E) + f_T(T; \mathbf{D}_T),
\end{equation}
where $f_E$ and $f_T$ are independent functions modeling the contributions of strong edges and texture distortions, respectively. This decoupled formulation is consistent with the observation that the perceptual contributions of edge degradation (on $\Omega_C$) and texture distortions (on $\Omega_H$) are perceived as separate phenomena.

The polynomial fitting framework is supported by the Taylor expansion hypothesis. By expanding $f(E, T; \mathbf{D})$ around a suitable reference point and assuming cross-sensitivity terms are negligible ($m, n \geq 1$), the second-order polynomial fitting reflects the dominant perceptual components of $E$ and $T$. This ensures that the resulting index $B \equiv \text{BELE}$ captures the primary perceptual variations while maintaining a balance between the contributions of strong edges and texture distortions.

In addition, the marginal metrics $f_E(E; \mathbf{D}_E)$ and $f_T(T; \mathbf{D}_T)$ can be considered affine functions of $E$ and $T$, respectively, provided their outputs exhibit a linear relationship with the corresponding subjective quality scores (DMOS). Under this condition, the approximation can be expressed as:
\begin{equation}
	B = D_0 + D_1^E E + D_1^T T,
\end{equation}
where the constant $D_0$ accounts for possible non-zero subjective quality scores DMOS assigned to the original images during experimental sessions, while $D_1^E$ and $D_1^T$ compensate for the different sensitivity of the subjective quality scores with respect to $E$ and $T$.

% *** blocco riutilizzabile in revisione ***
%\stfcomment{Per giustificare l'affermazione che le singole funzioni sono affini, possiamo far riferimento allo studio dettagliato del blur per l'indice E e possiamo creare uno scatter-plot per il solo rumore e vedere la linearità rispetto al secondo indice T (CPSNR, per qualsiasi dataset).}
% ******************************************

The affine combination assumes that edge degradation and texture distortions contribute additively to subjective quality, aligning with the previously introduced decoupling principle. This formulation preserves the independence of $E$ and $T$ while allowing their marginal effects to be linearly scaled.

The model includes five parameters: two for the edge-based index $E$ — $Q$ and $\tau$, which define its statistical anchor and viewing distance threshold — and three for the polynomial fitting process that integrates $E$ and $T$. The texture-based index $T$, derived from CPSNR, does not introduce additional parameters, as it is computed directly from the visual map data.

\section{Performance evaluation}
\label{sec:Performance evaluation}

% quintet classical and deep IQA datasets - figures and table
\begin{table*}[!t]
	\centering
	\begin{minipage}{\textwidth}
		\centering
		\begin{tabular}{@{}c@{}c@{}c@{}c@{}}
			\textbf{\tiny LIVE DBR2} & \textbf{\tiny TID2013} & \textbf{\tiny CSIQ} & \textbf{\tiny KADID-10K} \\
			\includegraphics[width=1.75in]{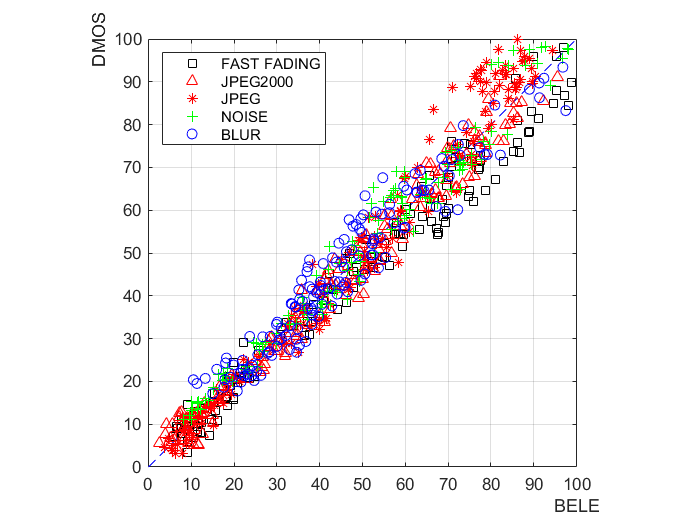} &
			\includegraphics[width=1.75in]{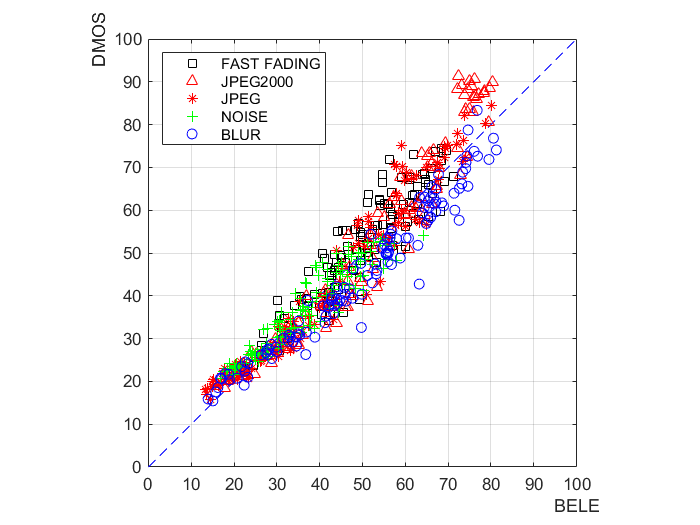} &
			\includegraphics[width=1.75in]{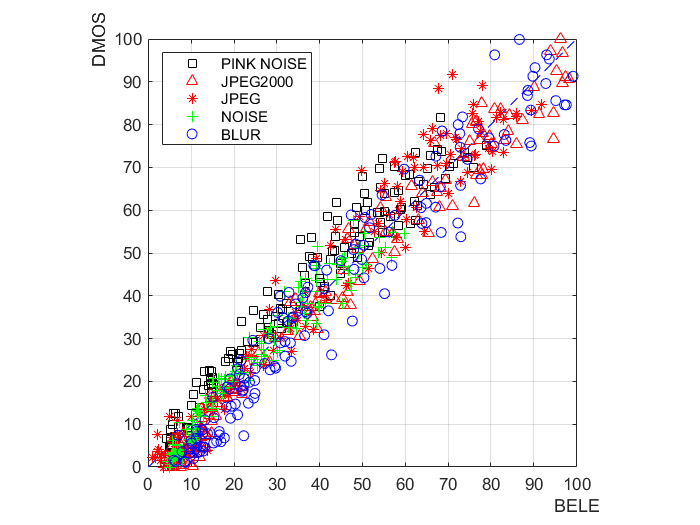} &
			\includegraphics[width=1.75in]{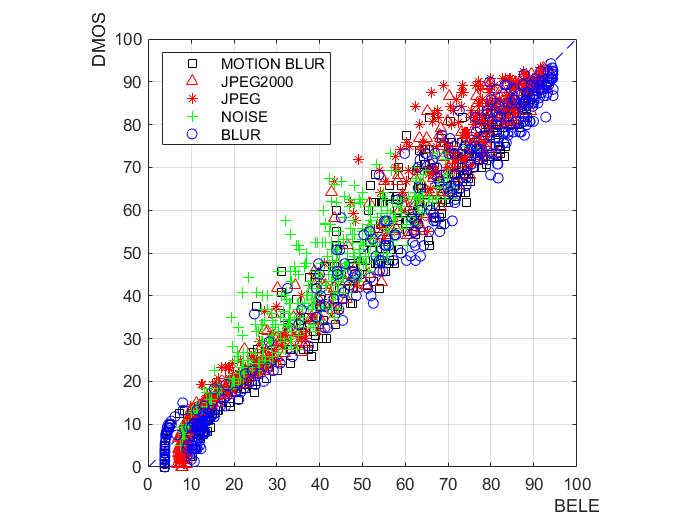} \\
		\end{tabular}
		\captionof{figure}{DMOS scatterplots compared to predicted values of BELE for various datasets and distortions. Columns represent datasets: LIVE DBR2 (1st column), TID2013 (2nd column), CSIQ (3rd column), and KADID-10K (4th column). Distortions common to all: Gaussian blur, white noise, JPEG, and JPEG2000; the fifth distortion varies: fast fading (LIVE DBR2, TID2013), pink noise (CSIQ), and motion blur (KADID-10K).}% Rows correspond to methods: BELE (top row), TOPIQ-FR (2nd row), and GMSD (bottom row).}
		\label{figure:IQA_quintet}
	\end{minipage}
	
	\vspace{0.5cm}
	
	\scalebox{0.85}{
		\begin{minipage}{1.15\textwidth}
			\centering
			\captionsetup{justification=justified, font=small}
			\caption{Comparison of BELE, DL-IQA, and classical IQA methods across five distortion subsets. Common distortions: Gaussian blur, white noise, JPEG, and JPEG2000. Fifth distortion: fast fading for LIVE DBR2 and TID2013, pink noise for CSIQ, motion blur for KADID-10K. Best values are bold blue, second best are bold black. DL-IQA methods require millions of parameters (M), while BELE and classical IQA methods use only five.}
			\label{table:IQA_quintet}
			\begin{tabular}{lccccccccccccc}
				\noalign{\smallskip}
				\toprule
				\multirow{2}{*}{IQA metric} & \multirow{2}{*}{Parameter} & \multicolumn{3}{c}{LIVE DBR2 quintet} & \multicolumn{3}{c}{TID2013 quintet} & \multicolumn{3}{c}{CSIQ quintet} & \multicolumn{3}{c}{KADID-10K quintet} \\
				\cmidrule(lr){3-5} \cmidrule(lr){6-8} \cmidrule(lr){9-11} \cmidrule(lr){12-14}
				& no. & RMSE & SROCC & LCC & RMSE & SROCC & LCC & RMSE & SROCC & LCC & RMSE & SROCC & LCC \\
				\midrule
				BELE & $\textcolor{blue}{\textbf{5}}$ & $\textcolor{blue}{\textbf{5.2033}}$ & $\textcolor{blue}{\textbf{0.98489}}$ & $\textcolor{blue}{\textbf{0.98170}}$ & $\textcolor{blue}{\textbf{5.3018}}$ & $\textcolor{blue}{\textbf{0.96592}}$ & $\textcolor{blue}{\textbf{0.95496}}$ & $\textcolor{blue}{\textbf{6.0803}}$ & $\textcolor{blue}{\textbf{0.97603}}$ & $\textcolor{blue}{\textbf{0.97483}}$ & $\textbf{5.5087}$ & $\textbf{0.98113}$ & $\textbf{0.98220}$ \\
				\midrule
				TOPIQ FR & $\scriptstyle\sim$ $35$ \text{M} & $\textbf{6.2515}$ & $\textbf{0.97590}$ & $\textbf{0.97347}$ & $\textbf{5.4183}$ & $\textbf{0.95519}$ & $\textbf{0.95291}$ & $\textbf{7.4203}$ & $\textbf{0.95922}$ & $\textbf{0.96210}$ & $\textcolor{blue}{\textbf{3.97223}}$ & $\textcolor{blue}{\textbf{0.98506}}$ & $\textcolor{blue}{\textbf{0.99079}}$ \\
				TOPIQ FR PIPAL & $\scriptstyle>$ $35$ \text{M} & $9.2458$ & $0.94319$ & $0.94100$ & $6.8388$ & $0.91356$ & $0.92384$ & $10.4726$ & $0.90639$ & $0.92298$ & $9.5824$ & $0.93110$ & $0.94515$ \\
				DISTS & $\scriptstyle\sim$ $12$ \text{M} & $8.9653$ & $0.94766$ & $0.94460$ & $9.1186$ & $0.85045$ & $0.85996$ & $11.0847$ & $0.90451$ & $0.91327$ & $12.2410$ & $0.90874$ & $0.90878$ \\
				LPIPS & $\scriptstyle\sim$ $12$ \text{M} & $10.9468$ & $0.92350$ & $0.91623$ & $8.7964$ & $0.86955$ & $0.87040$ & $12.0331$ & $0.90095$ & $0.89692$ & $13.3490$ & $0.89102$ & $0.89054$ \\
				LPIPS VGG & $\scriptstyle>$ $100$ \text{M} & $9.7321$ & $0.93185$ & $0.93441$ & $9.5204$ & $0.82956$ & $0.84620$ & $11.8233$ & $0.89198$ & $0.90068$ & $17.6900$ & $0.80402$ & $0.79772$ \\
				PIEAPP & $\scriptstyle\sim$ $80$ \text{M} & $11.6052$ & $0.91821$ & $0.90538$ & $8.7480$ & $0.89288$ & $0.87296$ & $11.7821$ & $0.91191$ & $0.90146$ & $15.5280$ & $0.93600$ & $0.86840$ \\
				\midrule
				VIF & $\textcolor{blue}{\textbf{5}}$ & $9.2403$ & $0.96359$ & $0.94107$ & $7.7535$ & $0.90342$ & $0.90093$ & $12.1935$ & $0.89068$ & $0.89399$ & $12.9868$ & $0.90607$ & $0.89667$ \\
				MS-SSIM & $\textcolor{blue}{\textbf{5}}$ & $8.7516$ & $0.95083$ & $0.94731$ & $7.2687$ & $0.91955$ & $0.91350$ & $9.6004$ & $0.93623$ & $0.93570$ & $13.2806$ & $0.89210$ & $0.89165$ \\
				FSIM & $\textcolor{blue}{\textbf{5}}$ & $7.6096$ & $0.96462$ & $0.96043$ & $6.0492$ & $0.95012$ & $0.94094$ & $8.9474$ & $0.94701$ & $0.94440$ & $11.3505$ & $0.91882$ & $0.92211$ \\
				GMSD & $\textcolor{blue}{\textbf{5}}$ & $7.6262$ & $0.96025$ & $0.96026$ & $6.2759$ & $0.94808$ & $0.93628$ & $8.5902$ & $0.95121$ & $0.94887$ & $9.6777$ & $0.93925$ & $0.94402$ \\
				\bottomrule
			\end{tabular}
		\end{minipage}
	}
\end{table*}

\noindent The VRF-based BELE method is calibration-free, allowing MOS/DMOS estimation across various viewing distances and applications without parameter optimization.  

To assess its performance, we statistically compare BELE with classical IQA methods and state-of-the-art deep learning-based approaches. The evaluation, conducted on identical datasets, focuses on linearity, absolute error, and computational cost, highlighting both the practical benefits and limitations of the proposed method.

The analysis first examines five common distortions — Gaussian blur, white noise, JPEG and JPEG2000 compression, and dataset-specific artifacts like fast fading, pink noise, or motion blur. This serves as a foundation for evaluating more complex mixed distortions, which pose greater challenges for traditional IQA methods. The comparison highlights predictive accuracy and robustness, demonstrating the benefits of separating edge and texture contributions in perceptual modeling.

% complete classical and deep IQA datasets - figures and table
\begin{table*}[!t]
	\centering
	\begin{minipage}{\textwidth}
		\centering
		\begin{tabular}{@{}c@{}c@{}c@{}c@{}}
			\textbf{\tiny TID2013} & \textbf{\tiny LIVE MD} & \textbf{\tiny KADID-10K} & \textbf{\tiny PIPAL} \\
			\includegraphics[width=1.75in]{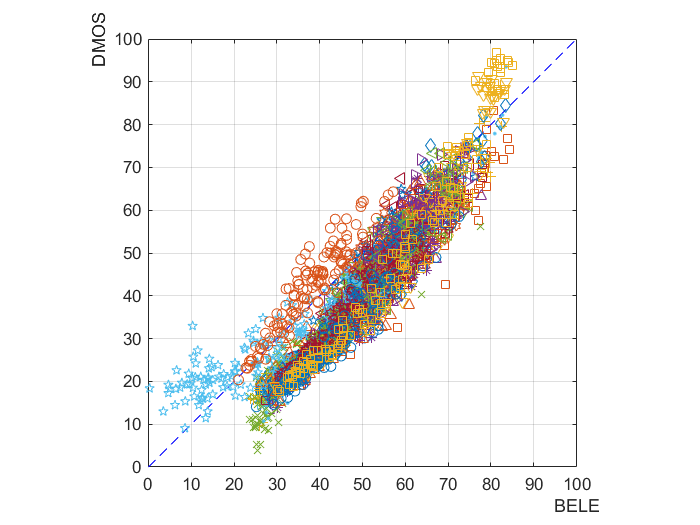} &
			\includegraphics[width=1.75in]{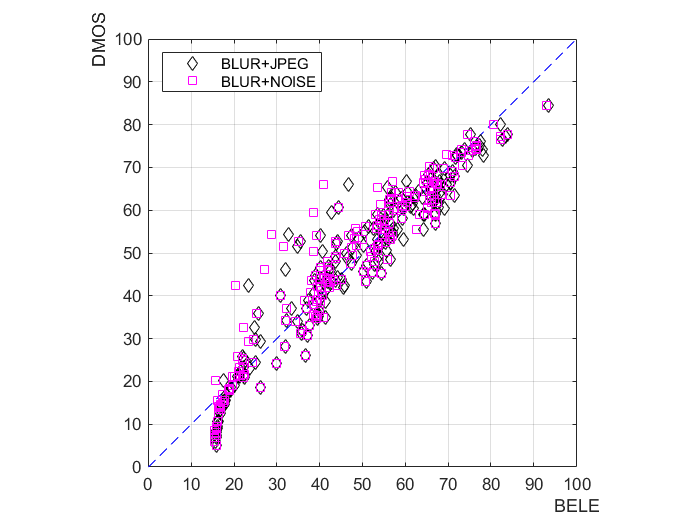} &
			\includegraphics[width=1.75in]{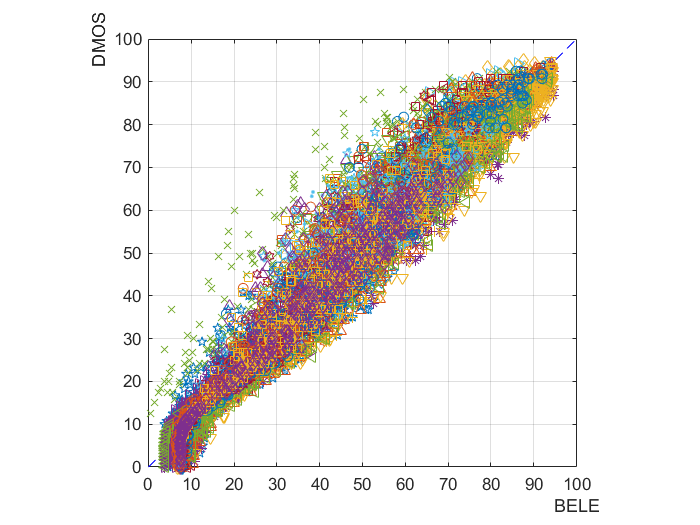} &
			\includegraphics[width=1.75in]{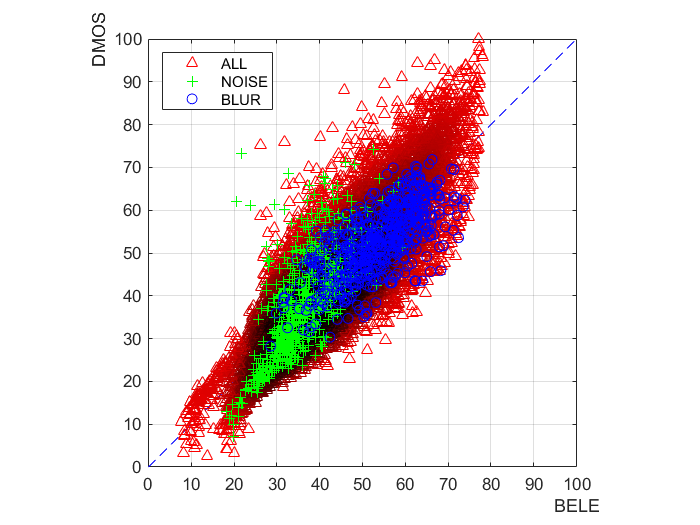} \\
		\end{tabular}
		\captionof{figure}{DMOS scatterplots compared to predicted values of BELE for various datasets and distortions. Columns represent datasets: TID2013 (1st column), LIVE MD (2nd column), KADID-10K (3rd column), and PIPAL (4th column). For all datasets, all distortions present in the datasets are shown. LIVE MD features combined distortions: Gaussian blur + JPEG and Gaussian blur + white noise. In PIPAL, blur (blue circles) and noise (green markers) are highlighted, with darker red areas indicating higher distortion density.}% Rows correspond to methods: BELE (top row), TOPIQ-FR (2nd row), and TOPIQ-FR-PIPAL (bottom row).}
		\label{figure:IQA_complete}
	\end{minipage}
	
	\vspace{0.5cm}
	
	\scalebox{0.85}{
		\begin{minipage}{1.15\textwidth}
			\centering
			\captionsetup{justification=justified, font=small}
			\caption{Comparison of BELE, DL-IQA, and classical IQA methods across entire datasets. Datasets include LIVE DBR2, TID2013, CSIQ, LIVE MD, KADID-10K, and PIPAL each containing a diverse set of distortions. Best values are highlighted in bold blue, and second best in bold black. DL-IQA methods require millions of parameters (M), while BELE and classical IQA methods use only five.}
			\label{table:IQA_complete}
			\begin{tabular}{lccccccccccccc}
				\noalign{\smallskip}
				\toprule
				\multirow{2}{*}{IQA metric} & \multirow{2}{*}{Parameter} & \multicolumn{2}{c}{LIVE DBR2} & \multicolumn{2}{c}{TID2013} & \multicolumn{2}{c}{CSIQ} & \multicolumn{2}{c}{LIVE MD} & \multicolumn{2}{c}{KADID-10K} & \multicolumn{2}{c}{PIPAL} \\
				\cmidrule(lr){3-4} \cmidrule(lr){5-6} \cmidrule(lr){7-8} \cmidrule(lr){9-10} \cmidrule(lr){11-12} \cmidrule(lr){13-14}
				& no. & SROCC & LCC & SROCC & LCC & SROCC & LCC & SROCC & LCC & SROCC & LCC & SROCC & LCC \\
				\midrule
				BELE & $\textcolor{blue}{\textbf{5}}$ & $\textcolor{blue}{\textbf{0.98489}}$ & $\textcolor{blue}{\textbf{0.98170}}$ & $\textcolor{blue}{\textbf{0.93718}}$ & $\textcolor{blue}{\textbf{0.91908}}$ & $\textbf{0.96410}$ & $\textbf{0.96027}$ & $\textcolor{blue}{\textbf{0.95165}}$ & $\textcolor{blue}{\textbf{0.95612}}$ & $\textbf{0.97031}$ & $\textbf{0.97034}$ & $\textcolor{blue}{\textbf{0.83845}}$ & $\textcolor{blue}{\textbf{0.84870}}$ \\
				\midrule
				TOPIQ FR & $\scriptstyle\sim$ $35$ \text{M} & $\textbf{0.97590}$ & $\textbf{0.97190}$ & $\textbf{0.91654}$ & $\textbf{0.91578}$ & $\textcolor{blue}{\textbf{0.96743}}$ & $\textcolor{blue}{\textbf{0.96971}}$ & $0.89093$ & $0.91437$ & $\textcolor{blue}{\textbf{0.98566}}$ & $\textcolor{blue}{\textbf{0.98515}}$ & $0.70902$ & $0.71528$ \\
				TOPIQ FR PIPAL & $\scriptstyle>$ $35$ \text{M} & $0.94319$ & $0.94100$ & $0.81986$ & $0.85481$ & $0.90756$ & $0.91860$ & $\textbf{0.89891}$ & $\textbf{0.92632}$ & $0.89474$ & $0.89443$ & $\textbf{0.81101}$ & $\textbf{0.84001}$ \\
				DISTS & $\scriptstyle\sim$ $12$ \text{M} & $0.94766$ & $0.94463$ & $0.70766$ & $0.75494$ & $0.92964$ & $0.93764$ & $0.78146$ & $0.81110$ & $0.81370$ & $0.81373$ & $0.58113$ & $0.59507$ \\
				LPIPS & $\scriptstyle\sim$ $12$ \text{M} & $0.92350$ & $0.91624$ & $0.74448$ & $0.77129$ & $0.92333$ & $0.91935$ & $0.74968$ & $0.82166$ & $0.82243$ & $0.81699$ & $0.58535$ & $0.58262$ \\
				LPIPS VGG & $\scriptstyle>$ $100$ \text{M} & $0.93185$ & $0.93361$ & $0.69395$ & $0.75864$ & $0.88304$ & $0.90424$ & $0.76774$ & $0.79994$ & $0.72005$ & $0.72837$ & $0.57315$ & $0.60666$ \\
				PIEAPP & $\scriptstyle\sim$ $80$ \text{M} & $0.91821$ & $0.91019$ & $0.84690$ & $0.83272$ & $0.89692$ & $0.88053$ & $0.80023$ & $0.86091$ & $0.86468$ & $0.77202$ & $0.70373$ & $0.69471$ \\
				\midrule
				VIF & $\textcolor{blue}{\textbf{5}}$ & $0.96359$ & $0.95876$ & $0.67697$ & $0.73360$ & $0.91936$ & $0.92515$ & $0.88271$ & $0.91857$ & $0.67919$ & $0.68563$ & $0.56008$ & $0.55826$ \\
				MS-SSIM & $\textcolor{blue}{\textbf{5}}$ & $0.95083$ & $0.94731$ & $0.78593$ & $0.83128$ & $0.91321$ & $0.89811$ & $0.81827$ & $0.85311$ & $0.82440$ & $0.75615$ & $0.55813$ & $0.58954$ \\
				FSIM & $\textcolor{blue}{\textbf{5}}$ & $0.96462$ & $0.96044$ & $0.85092$ & $0.87679$ & $0.93096$ & $0.91820$ & $0.86336$ & $0.90745$ & $0.85270$ & $0.79149$ & $0.58902$ & $0.60896$ \\
				GMSD & $\textcolor{blue}{\textbf{5}}$ & $0.96025$ & $0.96026$ & $0.80438$ & $0.85592$ & $0.95703$ & $0.92918$ & $0.84169$ & $0.88906$ & $0.84742$ & $0.80476$ & $0.58091$ & $0.62622$ \\
				\bottomrule
			\end{tabular}
		\end{minipage}
	}
\end{table*}

The evaluation employs four widely recognized datasets — LIVE DBR2 \cite{SHEIKH06B}, TID2013 \cite{PONOMARENKO15}, CSIQ \cite{LARSON10}, and KADID-10K \cite{LIN20} — selected for their diverse experimental protocols and DMOS calculation methodologies.

LIVE DBR2 follows a single stimulus protocol, where participants rate quality using a slider across five levels, with reference images included \cite{ITU12B}. TID2013 uses a tristimulus approach, where subjects compare two degraded images alongside the original reference. CSIQ adopts a multi-stimulus ranking method, displaying four images on an array of monitors. KADID-10K employs a single stimulus protocol with a hidden reference strategy.

%These varied methodologies introduce a broad spectrum of subjective quality assessments. If the proposed method performs consistently across all datasets, it would confirm its robustness and ability to capture fundamental aspects of subjective image quality, regardless of the specific evaluation protocol.

%Fig. \ref{figure:IQA_quintet} presents scatterplots comparing predicted quality scores with DMOS values across different datasets and distortions. Common distortions include Gaussian blur, white noise, JPEG, and JPEG2000, while the fifth distortion varies: fast fading (LIVE DBR2, TID2013), pink noise (CSIQ), and motion blur (KADID-10K). The scatterplots are organized by method: BELE (top row), TOPIQ-FR (second row), and GMSD (bottom row).
Fig. \ref{figure:IQA_quintet} presents scatterplots for BELE across these datasets and distortions. Common distortions include Gaussian blur, white noise, JPEG, and JPEG2000, while the fifth distortion varies: fast fading (LIVE DBR2, TID2013), pink noise (CSIQ), and motion blur (KADID-10K).

Tab. \ref{table:IQA_quintet} provides a quantitative summary of performance metrics: RMSE for average prediction error, SROCC for monotonicity, and PLCC for linearity in DMOS mapping. These metrics collectively assess the accuracy and robustness of each IQA method.

We compare our proposed method against several advanced IQA models. TOPIQ-FR \cite{CHEN24} employs a top-down approach with a Cross-Feature Attention Network (CFANet) based on ResNet50, integrating multi-scale features and cross-attention mechanisms. Trained on KADID-10K, it optimizes IQA performance with over 20 parameters. Additionally, we consider its PIPAL-trained variant (TOPIQ-FR-PIPAL) \cite{PIPAL20}, designed to handle diverse distortions and algorithmic corrections. 

DISTS \cite{DING22} assesses texture and structure similarity using a convolutional neural network, ensuring invariance to geometric transformations. It includes over 10 parameters, with $\alpha$ controlling structure-texture balance and $\beta$ normalizing similarity, allowing sensitivity adjustments for different distortions.

LPIPS \cite{ZHANG18} is evaluated in its AlexNet-based version and a deeper VGG-based variant (LPIPS-VGG), which enhances feature extraction for improved IQA performance. Lastly, PIEAPP \cite{PRASHNANI18} predicts subjective quality scores using perceptually calibrated loss functions, offering an alternative approach to IQA assessment.

The results show that BELE outperforms all methods across the tested datasets except for KADID-10K, where TOPIQ-FR performs best due to being specifically trained on this dataset. This highlights the reliance of learning-based models on their training data, especially for specific distortions or viewing conditions.

The comparison in Tab. \ref{table:IQA_quintet} also includes the number of parameters used by each method, demonstrating the simplicity of BELE, which remains as efficient as classical approaches. Despite its minimal parameterization, BELE achieves competitive or superior performance, confirming its robustness and ability to generalize across different datasets and distortion types.

% complete classical IQA datasets - table
\begin{table*}[!t]
%\begin{sidewaystable*}[htbp]
	%\centering
%	\scalebox{0.83}{
	\scalebox{1}{
	\begin{minipage}{\textwidth}
		\captionsetup{width=\textwidth}
		\caption{Experimental verification of BELE compared to DL-IQA methods on the full LIVE DBR2, TID2013, CSIQ, and LIVE MD datasets. Best values for each distortion are highlighted in bold blue, and second-best values in bold black. The table also includes computational cost (FLOPS) and the number of parameters used by each method. FLOPS refers to the computational cost per pair of processed images, with deep metric values expressed in gigaflops (GFLOPS). The number of parameters for deep metrics is expressed in millions (M).}
		\label{tab:classicIQAdatasetsComparison}
		
		\renewcommand{\arraystretch}{1.0} % Aumenta l'altezza delle righe della tabella
		\begin{adjustbox}{max width=1.0\textwidth}
			\begin{tabularx}{1.36\textwidth}{llcccccccccccccc}
				
				\toprule
				\multicolumn{2}{c}{} & \multicolumn{2}{c}{TOPIQ FR} & \multicolumn{2}{c}{TOPIQ FR PIPAL} & \multicolumn{2}{c}{DISTS} & \multicolumn{2}{c}{LPIPS} & \multicolumn{2}{c}{LPIPS VGG} & \multicolumn{2}{c}{PIEAPP} & \multicolumn{2}{c}{BELE} \\
				\midrule
				& Parameter no. & \multicolumn{2}{c}{$\scriptstyle\sim$ $35$ M} & \multicolumn{2}{c}{$\scriptstyle>$ $35$ M} & \multicolumn{2}{c}{$\scriptstyle\sim$ $12$ M} & \multicolumn{2}{c}{$\scriptstyle\sim$ $12$ M} & \multicolumn{2}{c}{$\scriptstyle>$ $100$ M} & \multicolumn{2}{c}{$\scriptstyle\sim$ $80$ M} & \multicolumn{2}{c}{\textcolor{blue}{\textbf{5}}} \\
				& GFLOPS & \multicolumn{2}{c}{$\scriptstyle\sim$ 19} & \multicolumn{2}{c}{$\scriptstyle>$ 19} & \multicolumn{2}{c}{$\scriptstyle\sim$ 62} & \multicolumn{2}{c}{$\scriptstyle\sim$ 62} & \multicolumn{2}{c}{$\scriptstyle>$ 62} & \multicolumn{2}{c}{$\scriptstyle\sim$ 155} & \multicolumn{2}{c}{$\textcolor{blue}{\textbf{$\sim 10^{-5}$}}$} \\
				
				\midrule
				Classical & \multirow{2}{*}{Distortion type} & \multirow{2}{*}{SROCC} & \multirow{2}{*}{LCC} & \multirow{2}{*}{SROCC} & \multirow{2}{*}{LCC} & \multirow{2}{*}{SROCC} & \multirow{2}{*}{LCC} & \multirow{2}{*}{SROCC} & \multirow{2}{*}{LCC} & \multirow{2}{*}{SROCC} & \multirow{2}{*}{LCC} & \multirow{2}{*}{SROCC} & \multirow{2}{*}{LCC} & \multirow{2}{*}{SROCC} & \multirow{2}{*}{LCC} \\
				datasets &  & \multicolumn{2}{c}{} & \multicolumn{2}{c}{} & \multicolumn{2}{c}{} & \multicolumn{2}{c}{} & \multicolumn{2}{c}{} & \multicolumn{2}{c}{} & \multicolumn{2}{c}{} \\
				\midrule
				\multirow{5}{*}{LIVE DBR2} & Gaussian Blur & \textcolor{blue}{\textbf{0.97574}} & \textcolor{blue}{\textbf{0.96764}} & 0.96495 & 0.95138 & \textbf{0.97117} & \textbf{0.97516} & 0.94480 & 0.90429 & 0.95011 & 0.95875 & 0.93058 & 0.89410 & 0.96600 & 0.96511 \\
				& Bit Errors in JPEG2000 Stream & \textbf{0.97028} & 0.96619 & 0.96466 & \textbf{0.96699} & 0.96019 & 0.95859 & 0.95333 & 0.95294 & 0.95849 & 0.96124 & 0.95158 & 0.94491 & \textcolor{blue}{\textbf{0.98648}} & \textcolor{blue}{\textbf{0.98852}} \\
				& JPEG Compression & \textbf{0.97361} & \textbf{0.98238} & 0.96391 & 0.96995 & 0.96826 & 0.96884 & 0.96338 & 0.96204 & 0.96602 & 0.96292 & 0.95280 & 0.95671 & \textcolor{blue}{\textbf{0.98323}} & \textcolor{blue}{\textbf{0.98740}} \\
				& JPEG2000 Compression & \textbf{0.97138} & \textbf{0.97659} & 0.94187 & 0.94676 & 0.95180 & 0.95603 & 0.94719 & 0.95235 & 0.93976 & 0.93687 & 0.94525 & 0.94842 & \textcolor{blue}{\textbf{0.98405}} & \textcolor{blue}{\textbf{0.98872}} \\
				& Gaussian White Noise & \textcolor{blue}{\textbf{0.98663}} & \textbf{0.97377} & 0.97189 & 0.96351 & 0.97394 & 0.95180 & 0.96993 & 0.93007 & 0.96847 & 0.96488 & 0.96912 & 0.94721 & \textbf{0.98524} & \textcolor{blue}{\textbf{0.98684}} \\
				\midrule
				\multirow{24}{*}{TID2013} & Colour Additive Noise & \textbf{0.85940} & \textbf{0.88616} & 0.67623 & 0.69979 & 0.77524 & 0.77573 & 0.71814 & 0.70401 & 0.65754 & 0.70349 & 0.72645 & 0.73543 & \textcolor{blue}{\textbf{0.97674}} & \textcolor{blue}{\textbf{0.96897}} \\
				& Gaussian Blur & \textbf{0.95781} & 0.94712 & 0.91187 & 0.91419 & 0.92792 & 0.91942 & 0.95482 & \textbf{0.94832} & 0.88136 & 0.87532 & 0.89515 & 0.88314 & \textcolor{blue}{\textbf{0.98110}} & \textcolor{blue}{\textbf{0.97463}} \\
				& Gaussian White Noise & \textbf{0.92780} & \textbf{0.92943} & 0.81106 & 0.79340 & 0.83561 & 0.82461 & 0.80732 & 0.77647 & 0.81112 & 0.79974 & 0.85045 & 0.83958 & \textcolor{blue}{\textbf{0.94456}} & \textcolor{blue}{\textbf{0.94168}} \\
				& High Frequency Noise & \textbf{0.90949} & \textbf{0.95531} & 0.82910 & 0.87718 & 0.86140 & 0.88995 & 0.83918 & 0.85558 & 0.84857 & 0.87994 & 0.87153 & 0.91126 & \textcolor{blue}{\textbf{0.97740}} & \textcolor{blue}{\textbf{0.96643}} \\
				& Impulse Noise & \textbf{0.81325} & \textbf{0.79520} & 0.60840 & 0.59069 & 0.67232 & 0.65150 & 0.66584 & 0.63559 & 0.62051 & 0.60636 & 0.76468 & 0.75597 & \textcolor{blue}{\textbf{0.87292}} & \textcolor{blue}{\textbf{0.87650}} \\
				& Masked Noise & 0.82655 & 0.84550 & 0.71230 & 0.78138 & 0.82612 & 0.85448 & 0.78751 & 0.82222 & 0.52466 & 0.68876 & \textbf{0.82928} & \textbf{0.85959} & \textcolor{blue}{\textbf{0.96144}} & \textcolor{blue}{\textbf{0.94085}} \\
				& Quantization Noise & \textbf{0.86733} & \textbf{0.87277} & 0.75364 & 0.74630 & 0.76881 & 0.76228 & 0.78646 & 0.77420 & 0.72886 & 0.73194 & 0.68225 & 0.66093 & \textcolor{blue}{\textbf{0.94374}} & \textcolor{blue}{\textbf{0.94280}} \\
				& Spatially Correlated Noise & \textcolor{blue}{\textbf{0.92012}} & \textcolor{blue}{\textbf{0.91890}} & 0.86254 & 0.85819 & 0.85465 & 0.83945 & 0.78417 & 0.76617 & 0.80961 & 0.80209 & 0.80005 & 0.80804 & \textbf{0.91737} & \textbf{0.91303} \\
				& Block-wise Distortions & 0.19702 & 0.30319 & 0.38817 & 0.46027 & 0.33489 & 0.30580 & 0.45282 & 0.51352 & \textbf{0.52651} & \textbf{0.54282} & 0.15881 & 0.17683 & \textcolor{blue}{\textbf{0.71846}} & \textcolor{blue}{\textbf{0.78224}} \\
				& Chromatic Aberrations & 0.89302 & 0.92203 & 0.86589 & 0.95021 & 0.88625 & 0.95555 & \textbf{0.89699} & 0.92945 & 0.83946 & \textbf{0.96440} & 0.87867 & 0.96144 & \textcolor{blue}{\textbf{0.96833}} & \textcolor{blue}{\textbf{0.97686}} \\
				& Comfort Noise & \textbf{0.93331} & \textcolor{blue}{\textbf{0.96060}} & 0.86779 & 0.91632 & 0.89312 & 0.89969 & 0.86773 & 0.86006 & 0.81385 & 0.90006 & 0.85927 & 0.91265 & \textcolor{blue}{\textbf{0.97052}} & \textbf{0.94717} \\
				& Contrast Change & 0.58866 & 0.74896 & 0.59626 & 0.72227 & 0.47560 & 0.70351 & 0.43942 & 0.54380 & 0.30372 & 0.37622 & \textbf{0.78930} & \textbf{0.85591} & \textcolor{blue}{\textbf{0.94185}} & \textcolor{blue}{\textbf{0.95566}} \\
				& Image Denoising & \textbf{0.94726} & \textbf{0.96406} & 0.89383 & 0.93530 & 0.89007 & 0.92538 & 0.88555 & 0.90359 & 0.85636 & 0.91105 & 0.84322 & 0.87811 & \textcolor{blue}{\textbf{0.98738}} & \textcolor{blue}{\textbf{0.98263}} \\
				& Dither Color Quantization & \textbf{0.91863} & \textbf{0.92083} & 0.80180 & 0.80875 & 0.81463 & 0.82240 & 0.79670 & 0.78004 & 0.79404 & 0.79746 & 0.88463 & 0.88149 & \textcolor{blue}{\textbf{0.97355}} & \textcolor{blue}{\textbf{0.96911}} \\
				& JPEG Compression & \textbf{0.92369} & \textbf{0.96303} & 0.88331 & 0.92695 & 0.88830 & 0.91495 & 0.89080 & 0.90589 & 0.87594 & 0.90679 & 0.84510 & 0.87099 & \textcolor{blue}{\textbf{0.97775}} & \textcolor{blue}{\textbf{0.97455}} \\
				& JPEG Transmission Errors & \textbf{0.91592} & \textbf{0.93322} & 0.84367 & 0.87318 & 0.91249 & 0.88573 & 0.90487 & 0.88561 & 0.71357 & 0.80341 & 0.85169 & 0.86723 & \textcolor{blue}{\textbf{0.93721}} & \textcolor{blue}{\textbf{0.93329}} \\
				& JPEG2000 Compression & \textbf{0.96366} & 0.94797 & 0.92882 & \textbf{0.95202} & 0.93089 & 0.94130 & 0.92545 & 0.93467 & 0.91506 & 0.93907 & 0.94272 & 0.94964 & \textcolor{blue}{\textbf{0.97827}} & \textcolor{blue}{\textbf{0.97179}} \\
				& JPEG2000 Transmission Errors & \textbf{0.90369} & \textbf{0.90648} & 0.88542 & 0.87749 & 0.86650 & 0.83445 & 0.81598 & 0.80886 & 0.78029 & 0.76516 & 0.85437 & 0.85859 & \textcolor{blue}{\textbf{0.91206}} & \textcolor{blue}{\textbf{0.91618}} \\
				& Lossy Compression & \textbf{0.94948} & \textbf{0.95371} & 0.91802 & 0.93107 & 0.92613 & 0.93125 & 0.90796 & 0.89874 & 0.90438 & 0.92035 & 0.86857 & 0.87879 & \textcolor{blue}{\textbf{0.95836}} & \textcolor{blue}{\textbf{0.95397}} \\
				& Mean Shift & \textbf{0.81506} & \textbf{0.82059} & 0.40003 & 0.46071 & 0.80293 & 0.80866 & 0.77628 & 0.80274 & 0.73048 & 0.69509 & 0.50405 & 0.49472 & \textcolor{blue}{\textbf{0.82935}} & \textcolor{blue}{\textbf{0.85472}} \\
				& Multiplicative Gaussian Noise & \textbf{0.89490} & \textbf{0.89799} & 0.76450 & 0.74778 & 0.78348 & 0.76232 & 0.72349 & 0.70283 & 0.74128 & 0.72747 & 0.82509 & 0.81782 & \textcolor{blue}{\textbf{0.95143}} & \textcolor{blue}{\textbf{0.94374}} \\
				& Non-Eccentricity Pattern Noise & 0.82772 & \textbf{0.88367} & 0.80788 & 0.84865 & \textbf{0.84411} & 0.86185 & 0.80460 & 0.80576 & 0.56508 & 0.48393 & 0.78298 & 0.79848 & \textcolor{blue}{\textbf{0.95789}} & \textcolor{blue}{\textbf{0.93972}} \\
				& Saturation Change & 0.80788 & \textbf{0.80052} & 0.47612 & 0.41774 & 0.79770 & 0.69480 & \textbf{0.81322} & 0.79161 & 0.60006 & 0.47802 & 0.69897 & 0.65795 & \textcolor{blue}{\textbf{0.91268}} & \textcolor{blue}{\textbf{0.93324}} \\
				& Sparse Sampling & \textbf{0.95972} & 0.94306 & 0.93447 & \textbf{0.95867} & 0.94066 & 0.94059 & 0.93815 & 0.94436 & 0.94094 & 0.95865 & 0.91400 & 0.93581 & \textcolor{blue}{\textbf{0.98494}} & \textcolor{blue}{\textbf{0.97043}} \\
				\midrule
				\multirow{6}{*}{CSIQ} & Gaussian White Noise & \textbf{0.96041} & \textbf{0.96201} & 0.86952 & 0.87738 & 0.92392 & 0.92232 & 0.92883 & 0.91415 & 0.92196 & 0.91569 & 0.94318 & 0.93977 & \textcolor{blue}{\textbf{0.98206}} & \textcolor{blue}{\textbf{0.97709}} \\
				& Gaussian Blur & \textbf{0.97200} & \textbf{0.97122} & 0.94857 & 0.96028 & 0.96023 & 0.96818 & 0.95002 & 0.93379 & 0.95368 & 0.95499 & 0.94827 & 0.92709 & \textcolor{blue}{\textbf{0.97629}} & \textcolor{blue}{\textbf{0.97457}} \\
				& JPEG Compression & 0.95119 & \textcolor{blue}{\textbf{0.98394}} & 0.92813 & 0.96363 & \textbf{0.96370} & 0.97777 & 0.95144 & 0.96838 & 0.95631 & 0.96676 & 0.95000 & 0.97522 & \textcolor{blue}{\textbf{0.96648}} & \textbf{0.97879} \\
				& Contrast Decrement & \textbf{0.95136} & \textbf{0.95489} & 0.92738 & 0.93159 & 0.94791 & 0.94228 & 0.94761 & 0.92431 & 0.91165 & 0.87544 & 0.94197 & 0.92855 & \textcolor{blue}{\textbf{0.95839}} & \textcolor{blue}{\textbf{0.96558}} \\
				& Additive Pink Gaussian Noise & \textbf{0.96488} & \textbf{0.96356} & 0.90333 & 0.90073 & 0.94172 & 0.93229 & 0.94522 & 0.93464 & 0.92539 & 0.91479 & 0.92559 & 0.91863 & \textcolor{blue}{\textbf{0.97437}} & \textcolor{blue}{\textbf{0.97321}} \\
				& JPEG2000 Compression & \textbf{0.96600} & \textbf{0.97579} & 0.92977 & 0.95319 & 0.95400 & 0.95908 & 0.93835 & 0.94964 & 0.95246 & 0.95506 & 0.95730 & 0.96737 & \textcolor{blue}{\textbf{0.98295}} & \textcolor{blue}{\textbf{0.98841}} \\
				\midrule
				\multirow{2}{*}{LIVE MD} & Blur + JPEG & 0.89785 & 0.92167 & \textbf{0.90259} & \textbf{0.92833} & 0.88377 & 0.88690 & 0.84737 & 0.89346 & 0.84184 & 0.85551 & 0.81801 & 0.87830 & \textcolor{blue}{\textbf{0.95582}} & \textcolor{blue}{\textbf{0.95731}} \\
				& Blur + Gaussian Noise & 0.89079 & 0.91116 & \textbf{0.89874} & \textbf{0.92696} & 0.79769 & 0.80668 & 0.76607 & 0.80842 & 0.76942 & 0.78241 & 0.78372 & 0.84699 & \textcolor{blue}{\textbf{0.94861}} & \textcolor{blue}{\textbf{0.95004}} \\
				\bottomrule
			\end{tabularx}
		\end{adjustbox}
	\end{minipage}
	}
\end{table*}
%\end{sidewaystable*}

% complete deep IQA datasets - table
\begin{table*}[!h]
	%\begin{sidewaystable*}[htbp]
	%\centering
%	\scalebox{0.83}{
	\scalebox{1}{
	\begin{minipage}{\textwidth}
		%\centering
		\captionsetup{width=\textwidth}
		\caption{Experimental verification of BELE compared to DL-IQA methods on the full KADID-10K and PIPAL datasets. Best values for each distortion are highlighted in bold blue, and second-best values in bold black. The table also includes computational cost (FLOPS) and the number of parameters used by each method. FLOPS refers to the computational cost per pair of processed images, with deep metric values expressed in gigaflops (GFLOPS). The number of parameters for deep metrics is expressed in millions (M).}
		\label{tab:deepIQAdatasetsComparison}
		
		\renewcommand{\arraystretch}{1.0} % Aumenta l'altezza delle righe della tabella
		\begin{adjustbox}{max width=1.0\textwidth}
			\begin{tabularx}{1.36\textwidth}{llcccccccccccccc}
				
				\toprule
				\multicolumn{2}{c}{} & \multicolumn{2}{c}{TOPIQ FR} & \multicolumn{2}{c}{TOPIQ FR PIPAL} & \multicolumn{2}{c}{DISTS} & \multicolumn{2}{c}{LPIPS} & \multicolumn{2}{c}{LPIPS VGG} & \multicolumn{2}{c}{PIEAPP} & \multicolumn{2}{c}{BELE} \\
				\midrule
				& Parameter no. & \multicolumn{2}{c}{$\scriptstyle\sim$ $35$ M} & \multicolumn{2}{c}{$\scriptstyle>$ $35$ M} & \multicolumn{2}{c}{$\scriptstyle\sim$ $12$ M} & \multicolumn{2}{c}{$\scriptstyle\sim$ $12$ M} & \multicolumn{2}{c}{$\scriptstyle>$ $100$ M} & \multicolumn{2}{c}{$\scriptstyle\sim$ $80$ M} & \multicolumn{2}{c}{\textcolor{blue}{\textbf{5}}} \\
				& GFLOPS & \multicolumn{2}{c}{$\scriptstyle\sim$ 19} & \multicolumn{2}{c}{$\scriptstyle>$ 19} & \multicolumn{2}{c}{$\scriptstyle\sim$ 62} & \multicolumn{2}{c}{$\scriptstyle\sim$ 62} & \multicolumn{2}{c}{$\scriptstyle>$ 62} & \multicolumn{2}{c}{$\scriptstyle\sim$ 155} & \multicolumn{2}{c}{$\textcolor{blue}{\textbf{$\sim 10^{-5}$}}$} \\
				
				\midrule
				Deep & \multirow{2}{*}{Distortion type} & \multirow{2}{*}{SROCC} & \multirow{2}{*}{LCC} & \multirow{2}{*}{SROCC} & \multirow{2}{*}{LCC} & \multirow{2}{*}{SROCC} & \multirow{2}{*}{LCC} & \multirow{2}{*}{SROCC} & \multirow{2}{*}{LCC} & \multirow{2}{*}{SROCC} & \multirow{2}{*}{LCC} & \multirow{2}{*}{SROCC} & \multirow{2}{*}{LCC} & \multirow{2}{*}{SROCC} & \multirow{2}{*}{LCC} \\
				datasets &  & \multicolumn{2}{c}{} & \multicolumn{2}{c}{} & \multicolumn{2}{c}{} & \multicolumn{2}{c}{} & \multicolumn{2}{c}{} & \multicolumn{2}{c}{} & \multicolumn{2}{c}{} \\
				\midrule
				\multirow{25}{*}{KADID-10K} & Brighten & \textbf{0.97537} & \textcolor{blue}{\textbf{0.98442}} & 0.86481 & 0.91864 & 0.94430 & 0.95998 & 0.94681 & 0.95647 & 0.92937 & 0.94727 & 0.92394 & 0.91189 & \textcolor{blue}{\textbf{0.98093}} & \textbf{0.98171} \\
				& Color Block & \textcolor{blue}{\textbf{0.80649}} & \textcolor{blue}{\textbf{0.84500}} & 0.45596 & 0.46950 & 0.59098 & 0.61858 & 0.56673 & 0.59153 & 0.58869 & 0.61251 & 0.28955 & 0.32018 & \textbf{0.71164} & \textbf{0.76615} \\
				& Color Diffusion & \textcolor{blue}{\textbf{0.97258}} & \textcolor{blue}{\textbf{0.99013}} & 0.88715 & 0.91636 & 0.85456 & 0.89611 & 0.88963 & 0.94401 & 0.89493 & 0.92047 & 0.81520 & 0.67359 & \textbf{0.94454} & \textbf{0.98378} \\
				& Color Quantization & \textbf{0.95666} & \textcolor{blue}{\textbf{0.96866}} & 0.82797 & 0.85427 & 0.79439 & 0.80767 & 0.74709 & 0.70835 & 0.71103 & 0.71953 & 0.84467 & 0.81905 & \textcolor{blue}{\textbf{0.96583}} & \textbf{0.95440} \\
				& Color Saturation HSV & \textcolor{blue}{\textbf{0.80888}} & \textcolor{blue}{\textbf{0.84352}} & 0.63272 & 0.63578 & 0.68471 & 0.69562 & 0.65939 & 0.67523 & 0.64714 & 0.60458 & 0.53231 & 0.51918 & \textbf{0.73812} & \textbf{0.76878} \\
				& Color Saturation Lab & \textcolor{blue}{\textbf{0.98286}} & \textcolor{blue}{\textbf{0.98875}} & 0.89979 & 0.90717 & 0.91801 & 0.94043 & 0.93341 & 0.93851 & 0.94112 & 0.95091 & 0.84740 & 0.81178 & \textbf{0.97064} & \textbf{0.98126} \\
				& Color Shift & \textcolor{blue}{\textbf{0.95843}} & \textcolor{blue}{\textbf{0.98073}} & 0.76389 & 0.89284 & 0.84300 & 0.84237 & 0.74589 & 0.79631 & 0.71597 & 0.79209 & 0.67999 & 0.77263 & \textbf{0.93431} & \textbf{0.97300} \\
				& Contrast Change & \textcolor{blue}{\textbf{0.93322}} & \textcolor{blue}{\textbf{0.92089}} & 0.60680 & 0.61621 & 0.79634 & 0.78406 & 0.79166 & 0.77819 & 0.73602 & 0.70958 & 0.44021 & 0.50022 & \textbf{0.90455} & \textbf{0.90171} \\
				& Darken & \textbf{0.94463} & \textcolor{blue}{\textbf{0.98303}} & 0.79460 & 0.93182 & 0.91098 & 0.95594 & 0.90264 & 0.95827 & 0.89488 & 0.95075 & 0.81972 & 0.87342 & \textcolor{blue}{\textbf{0.95557}} & \textbf{0.98138} \\
				& Denoise & \textcolor{blue}{\textbf{0.97546}} & \textcolor{blue}{\textbf{0.97578}} & \textbf{0.93123} & 0.93294 & 0.92626 & 0.90718 & 0.87658 & 0.86067 & 0.89648 & 0.89710 & 0.85894 & 0.79187 & 0.92845 & \textbf{0.95008} \\
				& Gaussian Blur & \textbf{0.97035} & \textcolor{blue}{\textbf{0.99384}} & 0.93870 & 0.97377 & 0.95948 & 0.96907 & 0.93648 & 0.96145 & 0.95449 & 0.96489 & 0.94452 & 0.90212 & \textcolor{blue}{\textbf{0.97949}} & \textbf{0.99019} \\
				& High Sharpen & \textcolor{blue}{\textbf{0.97983}} & \textcolor{blue}{\textbf{0.98335}} & 0.88798 & 0.88073 & 0.88042 & 0.84551 & 0.92158 & 0.91626 & 0.89331 & 0.85311 & 0.84753 & 0.78686 & \textbf{0.93622} & \textbf{0.94040} \\
				& Impulse Noise & \textcolor{blue}{\textbf{0.94489}} & \textcolor{blue}{\textbf{0.96037}} & 0.79529 & 0.82962 & 0.80790 & 0.82725 & 0.78047 & 0.79907 & 0.79799 & 0.81693 & \textbf{0.85688} & 0.85613 & 0.82889 & \textbf{0.87395} \\
				& JPEG & \textbf{0.96310} & \textcolor{blue}{\textbf{0.99158}} & 0.84557 & 0.95833 & 0.85927 & 0.95002 & 0.85583 & 0.91557 & 0.85837 & 0.93511 & 0.83424 & 0.91415 & \textcolor{blue}{\textbf{0.98349}} & \textbf{0.98877} \\
				& JPEG2000 & \textbf{0.96826} & \textcolor{blue}{\textbf{0.99184}} & 0.91138 & 0.95030 & 0.93404 & 0.92323 & 0.92503 & 0.90938 & 0.92568 & 0.91899 & 0.93086 & 0.90255 & \textcolor{blue}{\textbf{0.98210}} & \textbf{0.99096} \\
				& Jitter & \textcolor{blue}{\textbf{0.98300}} & \textcolor{blue}{\textbf{0.99226}} & 0.93006 & 0.96541 & 0.95556 & 0.96478 & 0.94839 & 0.95942 & 0.93958 & 0.93766 & 0.91831 & 0.87946 & \textbf{0.98235} & \textbf{0.98644} \\
				& Lens Blur & \textcolor{blue}{\textbf{0.97103}} & \textcolor{blue}{\textbf{0.98687}} & 0.91580 & 0.95293 & 0.93257 & 0.93177 & 0.84240 & 0.87335 & 0.90286 & 0.89019 & 0.87777 & 0.86559 & \textbf{0.95011} & \textbf{0.97802} \\
				& Mean Shift & \textbf{0.85247} & \textbf{0.88819} & 0.42712 & 0.49773 & 0.78466 & 0.78382 & 0.77140 & 0.68363 & 0.71711 & 0.69498 & 0.49188 & 0.46043 & \textcolor{blue}{\textbf{0.91340}} & \textcolor{blue}{\textbf{0.93504}} \\
				& Motion Blur & \textcolor{blue}{\textbf{0.98502}} & \textcolor{blue}{\textbf{0.99011}} & 0.93904 & 0.95720 & 0.95056 & 0.94138 & 0.91235 & 0.92623 & 0.93728 & 0.92417 & 0.93855 & 0.91153 & \textbf{0.97956} & \textbf{0.98442} \\
				& Multiplicative Noise & \textcolor{blue}{\textbf{0.98073}} & \textcolor{blue}{\textbf{0.98187}} & 0.89222 & 0.87379 & 0.87752 & 0.86844 & 0.87405 & 0.87095 & 0.84401 & 0.83089 & 0.92547 & 0.86637 & \textbf{0.95276} & \textbf{0.95994} \\
				& Non-Eccentricity Patch & \textbf{0.93829} & \textbf{0.94303} & 0.56757 & 0.54427 & 0.67234 & 0.64644 & 0.61386 & 0.59787 & 0.64602 & 0.58355 & 0.59739 & 0.62077 & \textcolor{blue}{\textbf{0.97933}} & \textcolor{blue}{\textbf{0.96335}} \\
				& Pixelate & \textcolor{blue}{\textbf{0.97074}} & \textcolor{blue}{\textbf{0.97474}} & 0.82077 & 0.87303 & 0.78431 & 0.84132 & 0.71828 & 0.74758 & 0.73193 & 0.78142 & 0.77939 & 0.75666 & \textbf{0.96172} & \textbf{0.96481} \\
				& Quantization & \textcolor{blue}{\textbf{0.96559}} & \textcolor{blue}{\textbf{0.96706}} & 0.74722 & 0.77789 & 0.80222 & 0.79325 & 0.84306 & 0.83683 & 0.78753 & 0.76141 & 0.77134 & 0.77060 & \textbf{0.91268} & \textbf{0.93324} \\
				& White Noise & \textcolor{blue}{\textbf{0.96575}} & \textcolor{blue}{\textbf{0.96577}} & 0.82003 & 0.80234 & 0.82011 & 0.81729 & 0.75739 & 0.76463 & 0.77774 & 0.78840 & 0.88115 & 0.83425 & \textbf{0.91565} & \textbf{0.92695} \\
				& White Noise Color Component & \textcolor{blue}{\textbf{0.97926}} & \textcolor{blue}{\textbf{0.98201}} & 0.88341 & 0.87821 & 0.88070 & 0.88341 & 0.84866 & 0.83976 & 0.85407 & 0.84869 & 0.91287 & 0.88107 & \textbf{0.94286} & \textbf{0.95528} \\
				\midrule
				\multirow{3}{*}{PIPAL} & Complete & 0.70902 & 0.71528 & \textbf{0.81101} & \textbf{0.84001} & 0.58113 & 0.59507 & 0.58535 & 0.58262 & 0.57315 & 0.60666 & 0.70373 & 0.69471 & \textcolor{blue}{\textbf{0.83845}} & \textcolor{blue}{\textbf{0.8487}} \\
				& Gaussian Blur & 0.61328 & 0.61826 & \textcolor{blue}{\textbf{0.74537}} & \textcolor{blue}{\textbf{0.74080}} & 0.48307 & 0.46940 & 0.48560 & 0.47462 & \textbf{0.73046} & \textbf{0.73995} & 0.69022 & 0.63949 & 0.70272 & 0.69958 \\
				& White Noise & 0.67082 & 0.66937 & \textcolor{blue}{\textbf{0.83173}} & \textcolor{blue}{\textbf{0.83859}} & \textbf{0.73207} & \textbf{0.73765} & 0.71404 & 0.71684 & 0.55618 & 0.51917 & 0.70489 & 0.70997 & 0.71494 & 0.70050 \\
				\bottomrule
			\end{tabularx}
		\end{adjustbox}
	\end{minipage}
	}
\end{table*}
%\end{sidewaystable*}

To ensure a comprehensive evaluation, we extend the analysis to all distortion types across multiple datasets, including LIVE MD \cite{ITU12B} and PIPAL \cite{PIPAL20}. LIVE MD employs a single stimulus methodology with a hidden reference approach, where participants rate distorted images without direct comparison to the original. In contrast, PIPAL uses a double stimulus side-by-side comparison, allowing direct quality assessment against reference images.

%Fig. \ref{figure:IQA_complete} shows scatterplots for BELE, TOPIQ-FR, and TOPIQ-FR-PIPAL across these datasets, while Tab. \ref{table:IQA_complete} summarizes their performance. BELE consistently ranks among the top performers, ranking second only in KADID-10K (where TOPIQ-FR was specifically trained) and in CSIQ, where its performance remains comparable to TOPIQ-FR.
Fig. \ref{figure:IQA_complete} shows scatterplots for BELE across these datasets, while Tab. \ref{table:IQA_complete} summarizes the performance compared to that of other metrics. BELE consistently ranks among the top performers, ranking second only in KADID-10K (where TOPIQ-FR was specifically trained) and in CSIQ, where its performance remains comparable to TOPIQ-FR.
%Tab. \ref{table:IQA_complete} shows the performance of BELE compared to that of other metrics across these datasets. BELE consistently ranks among the top performers, ranking second only in KADID-10K (where TOPIQ-FR was specifically trained) and in CSIQ, where its performance remains comparable to TOPIQ-FR.

For the PIPAL dataset, BELE surpasses TOPIQ-FR-PIPAL, achieving the best overall performance despite TOPIQ-FR-PIPAL being trained on this dataset. This highlights BELE's robustness and adaptability, even when compared to deep-learning-based methods tailored to specific datasets.

To further analyze performance across individual distortions, Table \ref{tab:classicIQAdatasetsComparison} presents results for classical datasets, while Table \ref{tab:deepIQAdatasetsComparison} focuses on deep-learning-based datasets.

For classical datasets, BELE outperforms competing methods for most distortions, demonstrating its effectiveness in handling traditional degradation types. In KADID-10K, BELE generally ranks second to TOPIQ-FR but remains competitive, highlighting its robustness despite the dataset being optimized for deep-learning-based approaches.

In PIPAL, BELE achieves the highest overall performance, even surpassing TOPIQ-FR-PIPAL in aggregate evaluations. However, for individual distortions such as Gaussian blur and white noise, BELE performs slightly worse than TOPIQ-FR-PIPAL.

These tables also include the calculation of FLOPS (Floating Point Operations Per Second) for each method, providing insights into computational efficiency. A detailed analysis of computational complexity is presented in Appendix \ref{app:Computational Complexity Analysis}.

\section{Conclusion}
\label{sec:Conclusion}

\noindent In this paper, we introduced BELE (Blur-Equivalent Linearized Estimator), a novel FR-IQA method that separates the perceptual effects of strong edge degradations and texture distortions into two distinct indices. The model employs a linearized estimator for blur degradations at varying viewing distances and uses CPSNR to characterize texture distortions. These indices are combined through polynomial fitting, resulting in a unified framework with only five parameters, making BELE both interpretable and computationally efficient.

The method was evaluated against classical and deep learning-based IQA approaches across multiple datasets, including LIVE DBR2, TID2013, CSIQ, KADID-10K, LIVE MD, and PIPAL. BELE consistently outperformed other methods on classical distortions, demonstrating strong generalization. While TOPIQ-FR performed best on the KADID-10K dataset due to targeted training, BELE remained competitive, particularly on datasets without such dataset-specific optimizations. However, BELE exhibited slightly lower performance for Gaussian blur and white noise on the PIPAL dataset, highlighting areas for potential improvement. Unlike deep-learning-based approaches, BELE achieves high accuracy without requiring extensive training or a large number of parameters, reinforcing its practical advantages.

This work underscores the value of interpretable, low-complexity models in FR-IQA and establishes BELE as a strong benchmark for diverse datasets. Future research could refine BELE’s handling of specific distortions and extend its applicability to Reduced-Reference (RR) and No-Reference (NR) IQA scenarios.

\begin{appendices}

\section{Computational Complexity Analysis}
\label{app:Computational Complexity Analysis}

% calcolo FLOPS
\noindent The computational complexity of BELE consists of two main components: the one-time precomputation of the spline interpolation to derive $\xi$ and the runtime evaluation for each image pair. Assuming input images of size $3 \times 224 \times 224$, the spline interpolation requires normalizing a dataset of size $N$, incurring $3N$ FLOPs, followed by polynomial evaluations, leading to a total cost of $3N + N \cdot (\log(S) + d + 1)$ FLOPs, where $d$ is the degree of the piecewise polynomial interpolation, with $S$ segments.

At runtime, BELE is computed for each image pair, requiring operations proportional to the number of pixels. If $C_{\text{BELE}}^{\text{pixel}}$ represents the FLOPs per pixel, the total runtime cost for $P$ image pairs is $P \cdot 224 \cdot 224 \cdot 3 \cdot C_{\text{BELE}}^{\text{pixel}}$. The overall FLOPs for BELE are thus:
\[
\text{FLOPs}_{\text{total}} = 3N + N \cdot (\log(S) + d + 1) + P \cdot 224 \cdot 224 \cdot 3 \cdot C_{\text{BELE}}^{\text{pixel}}.
\]

Additionally, methods requiring data rectification via logistic regression, such as VQEG, incur a significant computational cost. With a worst-case scenario of 250,000 iterations and $N = 1000$, the total cost is:
\[
\text{FLOPs}_{\text{logistic}} = (15N + 5) \cdot 250,000 = 3.75 \, \text{G}.
\]
Since BELE provides linearized outputs natively, it eliminates the need for this additional rectification, making it more computationally efficient.

\end{appendices}

\bibliographystyle{IEEEtran}
\bibliography{IEEEabrv,imageprocessing}

%\vskip 0pt plus -1fil
\newpage

\end{document}